\documentclass[preprint,amsmath,amssymb,3p, superscriptaddress,floatfix]{elsarticle}
\usepackage{amsmath}
\usepackage{amssymb}
\usepackage{natbib}
\biboptions{sort&compress}
\usepackage{graphicx}% Include figure files
\usepackage{dcolumn}% Align table columns on decimal point
\usepackage{bm}% bold math
\usepackage{hyperref}% add hypertext capabilities
\usepackage[T1]{fontenc}
\usepackage{float}%https://www.overleaf.com/project/5d516fba43ef842775448627https://www.overleaf.com/project/5d516fba43ef842775448627
\usepackage[normalem]{ulem}
\usepackage{color,xcolor}
\hypersetup{colorlinks=true,citecolor=blue,linkcolor=blue,filecolor=blue,
urlcolor=blue,pdfstartview=FitH,pdfpagemode=UseNone}
\usepackage[english]{babel}
\usepackage[thinlines]{easytable}

\usepackage{array}
\newcolumntype{P}[1]{>{\centering\arraybackslash}p{#1}}

\begin{document}

\title{Spatial mapping of disordered 2D systems: the conductance Sudoku}% Force line breaks with \\

\author{S. Mukim}  
\address{School of Physics, Trinity College Dublin, Dublin 2, Ireland}
\address{Centre for Research on Adaptive Nanostructures and Nanodevices (CRANN) \& Advanced Materials and Bioengineering Research (AMBER) Centre, Trinity College Dublin, Dublin 2, Ireland}
\author{C. Lewenkopf}
\address{Instituto de F\'{\i}sica, Universidade Federal Fluminense, 24210-346 Niteroi, Brazil}
\cortext[mycorrespondingauthor]{Corresponding author}
\ead{caio_lewenkopf@id.uff.br}

\author{M. S. Ferreira}
\address{School of Physics, Trinity College Dublin, Dublin 2, Ireland}
\address{Centre for Research on Adaptive Nanostructures and Nanodevices (CRANN) \& Advanced Materials and Bioengineering Research (AMBER) Centre, Trinity College Dublin, Dublin 2, Ireland}

\date{\today}

\begin{abstract}
 Motivated by recent advances on local conductance measurement techniques at the nanoscale, timely questions are being raised about what possible information can be extracted from a disordered graphene sheet by selectively interrogating its transport properties. Here we demonstrate how an inversion technique originally developed to identify the number of scatterers in a quantum device can be adapted to a multi-terminal setup in order to provide detailed information about the spatial distribution of impurities on the surface of graphene, as well as other 2D material systems. The methodology input are conductance readings (for instance, as a function of the chemical potential) between different electrode pairs, the output being the spatially resolved impurity density. 
 We discuss how the obtained spatial resolution depends on the number of such readings and on the electrode  geometry.
 Furthermore, by separating the impurity locations into partitions arranged in a grid-like geometry, this inversion procedure resembles a Sudoku puzzle in which the compositions of different sectors of a device are found by imposing that they must add up to specific constrained values established for the grid rows and columns. We argue that this technique may provide alternative new ways of extracting information from a disordered material through the selective probing of local quantities. 
\end{abstract}

\maketitle

\section{Introduction}

Inverse Problems (IP) in science are those that attempt to obtain from a set of observations the causal factors that generated them in the first place. IP are intrinsic parts of several visualization tools \cite{medical, fwi, tromp2008spectral, sonar} but are far less ubiquitous in the quantum realm \cite{Lassas2008,PhysRevLett.111.090403}. 
%\sout{In fact, the literature on the field of Quantum IP (QIP) is centred heavily on the fundamentals of inversion processes, e.g. whether a problem is ill-posed \cite{Lassas2008}, whether solutions are unique, stable \cite{PhysRevLett.111.090403}, etc.} 
In contrast, Materials Science appears as fertile ground for applications of IP since it involves studying the physical properties of structures for which the underlying Hamiltonians are often unknown \cite{PhysRevX.8.031029,Lai_2017,tsymbal,jasper,gianluca,
%PhysRevLett.97.046401,
Franceschetti1999,liping,plasma}.

Finding the precise parent Hamiltonian that generates a specific observable is an arduous process. In general, it consists of solving the Schr\"odinger equation with a Hamiltonian containing one (or more) variable parameter(s) which are varied until the solution closely matches the original observation. Because of the enormous number of possibilities, finding the exact configuration may be a computationally demanding task. It is worth stating that efficient codes and powerful computers alone are not sufficient to make this approach feasible and alternative ways of probing the phase space of possibilities are required. Neural-network-based search engines \cite{jensen,yazyev}, genetic algorithms \cite{Zhang2013} and, more recently, machine-learning strategies \cite{Vargas2019,kyriienko,anatole,fazli,burak,collins,Xia2018,dral,melko, Fazzio2019} have been proposed and, with different degrees of success, can speed up the search for the "inverted" configuration. As a result, simulations are starting to have an impact in reducing the time and cost associated with materials design, specially those involving high throughput studies of material groups \cite{Ziletti2018,rajan,suram,Koinuma2004,choi,nardelli,PhysRevLett.108.068701,
% Yan2015,
Fischer2006,Gautier2015}. These are large scale simulations that generate a gigantic volume of data with the intention of identifying optimal combinations that can be subsequently used as candidates for an exploratory experimental search of new materials. Despite the advances offered by these approaches, so far they are all based on “black-box” algorithms that bring little insight into the problems they aim to address \cite{Schmidt2019}.
Moreover, it should be stressed that disorder effects have not been addressed by this line of research.

Disorder is ubiquitous in graphene systems. As a consequence, due to quantum interference their ballistic and diffusive electronic transport properties show universal fluctuations at low temperatures \cite{Lewenkopf2010, DasSarma2011}.
Although the average conductance of these systems can be understood in terms of simple quantities, such as the kind of defects and their concentration, quite often the fluctuations make the comparison between theory and experiment particularly challenging.
A recent communication \cite{shardul} proposes a procedure to accurately identify the average conductance from a given finite data set (with fluctuations) and, further, puts forward an inversion technique that uses this information to extract structural and compositional information from a disordered graphene device.
Using the energy-dependent two-terminal conductance as the sole input, the reported inversion method identifies the exact number of impurities within the disordered material. Furthermore, it is particularly suitable to be used in carbon based, as well as in other 2D materials, is very stable and works in the ballistic, diffusive as well as  
at the onset of the localised transport regimes. While the ability to determine the precise number of impurities in a device through conductance inversion is in itself a significant achievement, the referred technique is not able to capture the spatial distribution of impurities but only the average concentration between a pair of electrodes. In this manuscript, we demonstrate 
that the inversion technique can be used to handle a multi-terminal disordered device. Interestingly, this generalisation gives enormous flexibility to how a disordered structure can be interrogated and, in doing so, leads to an inversion tool that obtains spatial information about the impurity concentration, something that Ref.~\cite{shardul} lacked. Remarkably, by
resolving the impurity distributions into system partitions, that we call ``cells", 
arranged in a grid-like geometry, the inversion procedure resembles a Sudoku puzzle in which the compositions of different sectors of a device are obtained by imposing that they must add up to specific values established for the grid rows and columns.

The experimental setup required to implement our proposal can be easily realised with current nanoscale fabrication techniques. 
Multi-terminal experiments (see, {\it e.g}, Ref.~\cite{Ihn2010} for an introduction) have played a key role in the understanding of the electronic transport properties of graphene \cite{Ghahari2011, Perkins2013, doi:10.1021/acsnano.6b05288}, as well as in a variety of 2D systems \cite{Cui2015,SHIN201959} and surface states in topological insulators \cite{Perkins2013}. 
Through multiple measurements of local and non-local resistances \cite{Buttiker1986}, one can in principle experimentally determine the multiterminal conductance matrix. The possibilities have been vastly improved by techniques that enable to directly probe the local conductances at the nanoscale, some of them already applied to carbon based systems, such as scanning gate microscopy \cite{Jura2007, Braem2018, Brun2019, Brun2020} and multi-probe scanning tunneling spectroscopy \cite{Ping2013, Baringhaus2014}, to name but a few. 
These recent experimental developments provide extra motivation to explore what type of information can be obtained from a disordered device through selectively interrogating its transport properties, which is precisely the underlying idea behind this work.   

This manuscript is structured as follows: We begin presenting the inversion methodology in a multi-electrode framework and applying it first to an illustrative case in which the impurity concentration is captured as a whole but not spatially resolved. Subsequently, we demonstrate how the same inversion tool can be implemented to extract  information about the spatial distribution of impurities from the conductance between terminals, sampling the electronic transport flow at different parts of the device. This will be done for a few different cases where the device is divided into a number of cells that determines the spatial resolution of the inversion. 

%------------------------------------------------------------------------------------
\section{Model and Methods}

\begin{figure}[!h]
    \centering
    \includegraphics[width=0.350\columnwidth]{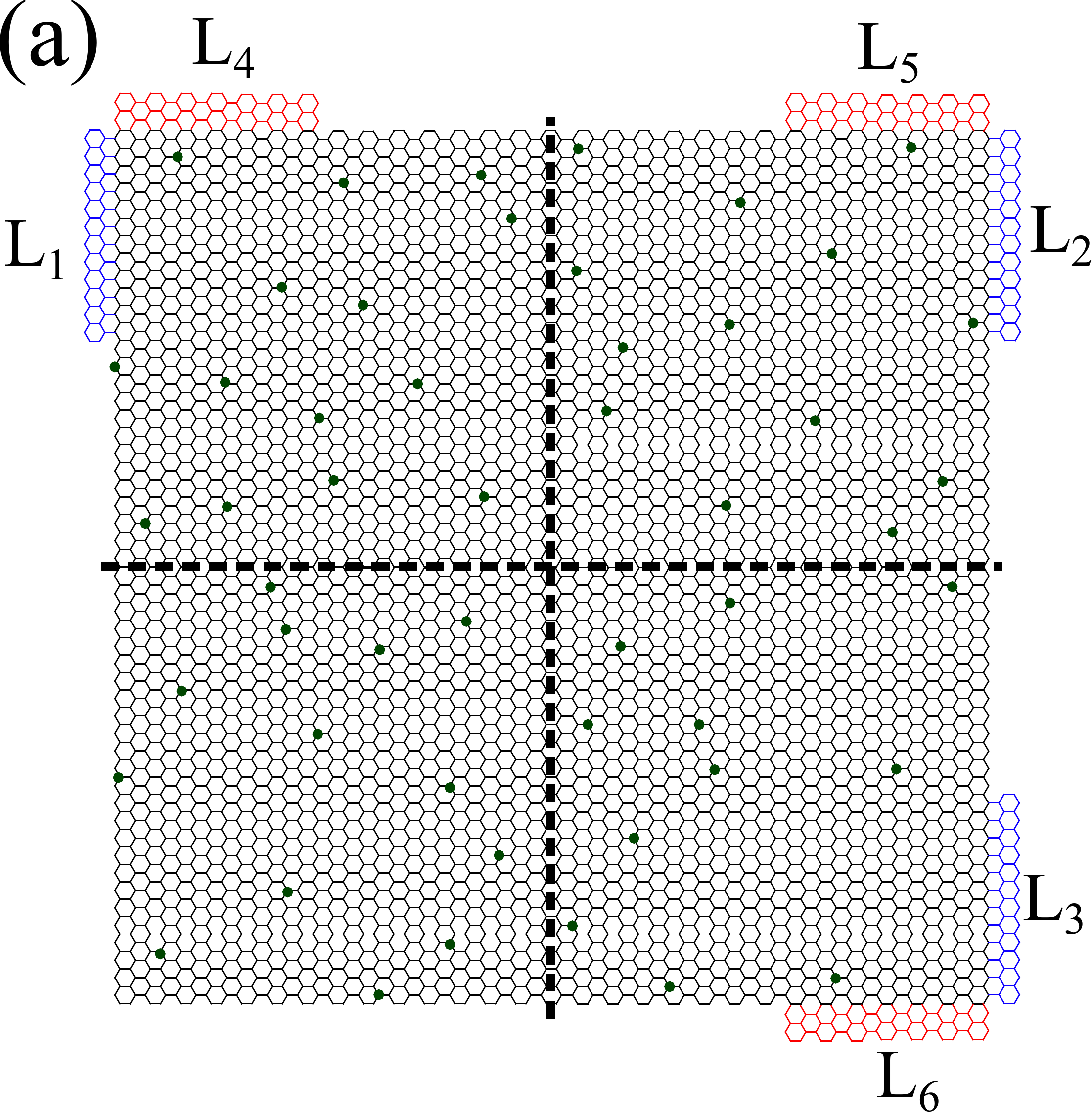}
    \includegraphics[width=0.47\columnwidth]{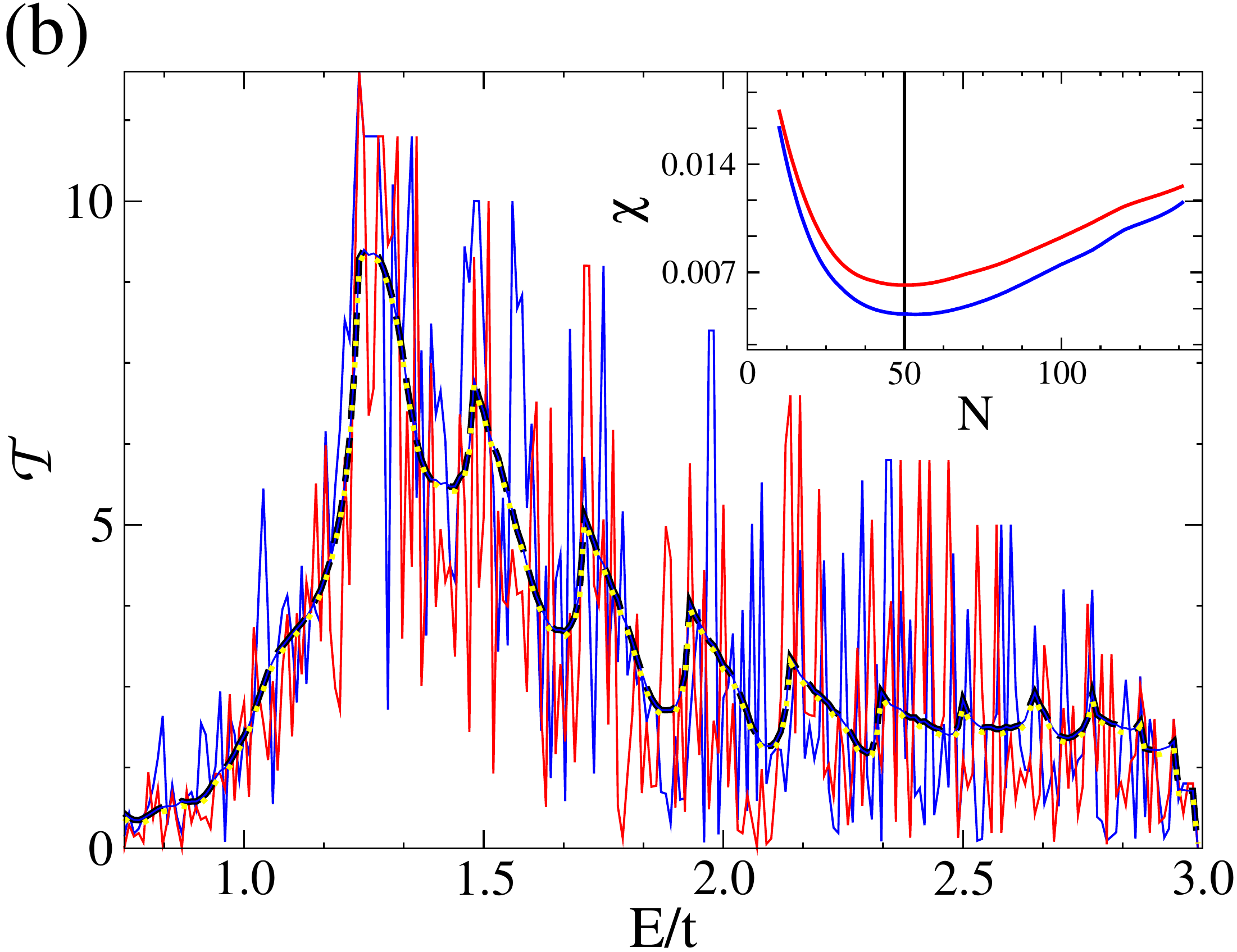}
    \caption{(a) Schematic representation of a graphene sheet contacted to multiple electrodes. Electrodes are depicted as short coloured extensions of the honeycomb structure, labelled $L_1$ to $L_6$.
    %\sout{that span the entire device length. Vertical (horizontal) electrodes are labelled as $L_1$ and $L_2$ ($L_3$ and $L_4$).} 
    The device possesses a number of randomly distributed impurities represented by solid dots. Dashed lines serve to delineate some of the cells subsequently used to establish the spatial distribution of impurities.  (b) Transmissions ${\cal T}_{1,2}$ (blue) and ${\cal T}_{4,6}$ (red) for a graphene sheet containing a total of $50$ impurities, which translates into an overall concentration of $n=1.5\%$, both plotted as a function of energy (in units of the electronic hopping integral $t$). Thick dashed line represents the CA transmissions $\langle {\cal T}_{1,2}\rangle$ for the same impurity concentration. $\langle {\cal T}_{4,6}\rangle$ is not shown to avoid too congested a figure but it is very similar to $\langle {\cal T}_{1,2}\rangle$. The inset plots the concentration-dependent misfit functions $\chi_{1,2}(N)$ and $\chi_{4,6}(N)$. }
    \label{fig1}
\end{figure}

Fig.~\ref{fig1}(a) shows a schematic diagram of the multi-terminal setup considered here and will be used as a reference throughout the manuscript. It consists of a graphene system connected to multiple electrodes represented by the coloured stripes labelled $L_1$ to $L_6$ located on the edges of the device.
Horizontal (Vertical) edges are chosen to be along the crystallographic armchair (zigzag) direction. We stress that since the terminals are placed on both kinds of edge terminations, this arbitrary choice of boundary conditions has no impact on any of our findings. The graphene sheet contains a finite number of randomly distributed impurities represented by  dots, here assumed to represent substitutional atoms other than those of the host 2D material. Adsorbed impurities \cite{Duffy2016} or vacancies \cite{Ridolfi2017} can be easily considered as well. In this paper, due to its relevance and model simplicity \cite{neto2009electronic}, we choose a graphene sheet as a case in point. We stress that the inversion methodology is also applicable to a variety of carbon-based materials and 2D systems, see for instance Ref.~\cite{Duarte_2021}.

Our analysis employs the Landauer-B\"uttiker approach to describe the multi-terminal electronic transport properties. The latter has been very successfully used in both ballistic and diffusive graphene systems \cite{Lewenkopf2010, DasSarma2011}.
In linear response theory, the multi-terminal Landauer-B\"uttiker formula for the electronic current $I_j$ at the terminal $j$ reads \cite{Buttiker1986,Ihn2010,Lima2018,Lima2021}
\begin{align}
I_j = \sum_{\ell=1}^{N} \mathcal G_{j,\ell} \left(V_j-V_\ell \right),
\label{current}
\end{align}
where $V_j$ is the voltage applied to the $j$-terminal and ${\cal G}_{j,\ell} $
is the conductance given by
\begin{align}
\mathcal G_{j,\ell} = \frac{2e^2}{h}\int_{-\infty}^{\infty} dE \left(-\frac{\partial f}{\partial E} \right) {\cal T}_{j,\ell}(E)
\label{eq:conductance}
\end{align}
expressed in terms of the the Fermi distribution  $f(E) = [1 + e^{(E-\mu)/k_B T}]^{-1}$ and the transmission coefficient ${\cal T}_{j,\ell} (E)$. The factor $2$ assumes spin degeneracy. 
The conductance $\mathcal{G}_{j,\ell}$ is a function of the chemical potential $\mu$ and, in principle, can be either calculated or measured experimentally (see discussion below). For simplicity, we use the transmission coefficients ${\cal T}_{j,\ell} (E)$ in our analysis.

The transmission ${\cal T}_{j,\ell} (E)$ is given by \cite{MeirWingreen1992}
\begin{align}
\label{eq:transmission}
{\cal T}_{j,\ell} (E)= {\rm tr} \left[ \mathbf \Gamma_\ell(E) \mathbf G^r(E) \mathbf \Gamma_j(E) \mathbf G^a(E) \right]
\end{align}
where $\mathbf G^r=\left(\mathbf G^a\right)^\dagger$ is the retarded Green's function of the full system,
while $\mathbf \Gamma_\ell$ is the line or decay width matrix of the lead corresponding to the $\ell$-terminal.
Both $\mathbf G^r$ and $\mathbf \Gamma_\ell$  are conveniently expressed in a discrete representation, typically a set of Wannier-like basis states. While $\mathbf G^r$ has the dimension of the number of Wannier-like states in the central region, the dimension of $\mathbf \Gamma_\ell$ is the number of states at the $\ell$-lead-central region interface.
The leads are considered as semi-infinite and the decay widths are computed by the standard prescription \cite{lewenkopf2013recursive}. Note that the most significant contribution to the integral in Eq.~\eqref{eq:conductance} comes from the transmission ${\cal T}_{j,\ell}(E)$ at energies close to the chemical potential (Fermi level) $\mu$. For that reason, throughout this manuscript we shall use the dimensionless energy-dependent transmission coefficient ${\cal T}_{j,\ell}$ as a proxy for the conductance between electrodes $j$ and $\ell$.

Multi-probe experiments measure the resistance $R_{j, \ell}$ and non-local voltages $V^{\rm NL}_{j',\ell'}$ between different pairs of electrodes (see, for instance, Ref.~\cite{Ihn2010}). Using gauge invariance and conservation of current, B\"uttiker \cite{Buttiker1986} has shown that $R$ and $V^{\rm NL}$ can be cast in terms of the conductance matrix.  Hence, by taking different combinations of biased electrodes and voltage probes, one can reconstruct ${\cal G}_{j,\ell}$ from experimentally obtained  $R_{j,\ell}$ and $V^{\rm NL}_{j,\ell}$ data.
It is important to mention that the conductance ${\cal G}_{j,\ell}$ depends on the system-electrode contact resistances, but the multi-probe set up allows one to determine them and extract the system electronic properties \cite{Buttiker1986}. 

The multi-terminal conductance expression, Eq.~\eqref{eq:conductance}, is written in terms of Green functions and as such it is model independent. In other words, once the Hamiltonian is known one can find the corresponding Green function and subsequently obtain the energy dependent conductance of the system. 
%\soutcaio{For the sake of simplicity, we describe the electronic structure of the material within the tight-binding model. Once again, this is in no way a requirement of the inversion method, which has been proven to work also with other electronic structure models regardless of the number of atomic orbitals involved \cite{shardul, fabio}.}
The inversion method has been shown to work for single- and multiple-orbital \cite{Duarte_2021} tight-binding models, as well as for density functional theory calculations \cite{shardul}. We stress that the method does not rely on any specific property of the electronic structure calculation implementation.

For the sake of simplicity, we model the graphene electronic properties using a single-orbital nearest neighbour tight-binding 
Hamiltonian \cite{neto2009electronic}, namely 
$H = -t \sum_{\langle i,j \rangle} (c^\dagger_i c^{}_j + \rm{H.c.})$, where $t$ is the hopping 
integral that we adopt as the energy unit and $c^\dagger_i \, (c^{}_i)$ stands for the operator of electron creation (annihilation) at the $i^{th}$ lattice site, and $\langle \cdots \rangle$ indicates that the sum is restricted over pairs of nearest neighbor sites. The disorder is modeled by adding to the Hamiltonian $H$ the local potential term
$U = \lambda \sum_i  c^\dagger_i c^{}_i$, where $i$ runs over the $N$ 
randomly-selected impurity sites and $\lambda = t$ is the impurity on-site energy. The number of impurities $N$ and its corresponding concentration $n$ may be used interchangeably in the manuscript.

The inversion procedure introduced in \cite{shardul} allows one to obtain $N$ from transport measurements by combining Configuration-Average (CA) with the ergodic principle that associates average (energy) conductance of a single sample with sample-to-sample average 
at a fixed energy. More specifically, the ergodic hypothesis assumes that
a running average over a continuous parameter upon which the conductance depends is equivalent to the ensemble average over different impurity configurations. In mathematical terms, this is done by considering the so-called misfit function defined as %
\begin{equation}
\chi_{j,j^\prime}(N)=\frac{1}{{{\cal E}_+}-{{\cal E}_-}} \, \int_{{\cal E}_-}^{{\cal E}_+}\! dE   \big[{\cal T}_{j,j^\prime}(E) - \langle {\cal T}_{j,j^\prime}(E,N) \rangle\big]^2\,,
\label{misfit}
\end{equation}
where the integration limits ${\cal E}_+$ and ${\cal E}_-$ are arbitrarily chosen energy values within the conduction band, provided ${\cal E}_+-{\cal E}_-$ is much larger than the transmission autocorrelation width.
The CA transmission $\langle {\cal T}_{j,j^\prime} \rangle$ is defined as $\langle {\cal T}_{j,j^\prime} \rangle = (1/M)\sum_{m=1}^M {\cal T}_{j,j^\prime}^{(m)}$, where the superscript $(m)$ labels the different realisations of disorder configurations.
While both ${\cal T}_{j,j^\prime}$ and $\langle {\cal T}_{j,j^\prime} \rangle$ are functions of energy, the latter is also a function of impurity concentration $n$. When plotted as a function of $N$ 
(or $n$), the misfit function is expected to display a 
minimum at a value that corresponds to the actual impurity concentration. 
It is straightforward to generalise the misfit function to resolve the impurity distribution on a grid with $N_{\rm grid}$ partitions, namely, $\chi_{j,j^\prime}(\{N_\nu\})$ with $\nu = 1, \cdots, N_{\rm grid}$. However, the inversion problem would involve finding the global minimum of a $N_{\rm grid}$-dimensional function, which is a quite challenging task for increasing $N_{\rm grid}$.
In what follows we discuss and show the results of a more practical and ingenious inversion procedure based on accounting for few constraints.

%------------------- FIG. 1 ---------------

\section{Results}

Figure \ref{fig1}(b) shows the calculated values of ${\cal T}_{1,2}$ and ${\cal T}_{4,6}$ of a 
device of size $D_x=30\sqrt{3} a$ and $D_y=50a$, $a$ being the lattice parameter of graphene, containing a total of $N=50$ substitutional impurities, {\it i.e.} a concentration of $n \approx 1.5\%$, both plotted as a function of energy. 
Although the actual number of impurities and how they are spatially distributed in the parent Hamiltonian are needed to generate ${\cal T}_{1,2}$ and ${\cal T}_{4,6}$, this information is not to be used in any part of the subsequent calculations but serves only as a reference to be compared against the final inverted results. Therefore, the conductance 
is the sole input function 
for the inversion procedure. 
The transmissions ${\cal T}_{1,2}$ and ${\cal T}_{4,6}$ show fluctuations of order unit over small energy scales, consistent with the universal conductance fluctuation characteristic of (chaotic) ballistic and diffusive electronic transport \cite{lee1985universal, Beenakker1997}. Hence, as previously alluded to, a naive approach attempting a comparison of the input function with the conductance for every single possible configuration is neither practical, since involves a prohibitively large number of combinations, nor insightful.

Fig.~\ref{fig1}(b) also displays the misfit functions $\chi_{1,2}$ and $\chi_{4,6}$ plotted as a function of the impurity number $N$. Both curves are obviously different but, reassuringly, possess distinctive minima at the same concentration value of $N=50$, which coincides exactly with the  chosen impurity concentration that generated the transmission matrix elements ${\cal T}_{ij}$. Furthermore, when plotting the CA transmissions 
$\langle {\cal T}_{1,2} \rangle$ and $\langle {\cal T}_{4,6} \rangle$ evaluated for the same impurity concentration we find very good agreement with the input functions $ {\cal T}_{1,2}$ and ${\cal T}_{4,6}$ despite the fluctuations, as seen in Fig. \ref{fig1}(b). This is unmistakable evidence that the inversion method introduced in \cite{shardul} can indeed account for the case of multiple electrodes.

%\sout{Figure \ref{fig1}(b) also shows the CA conductances $\langle {\cal T}_{12} \rangle$ and $\langle {\cal T}_{34} \rangle$ evaluated for the impurity concentration of $n=1.5\%$. 
%\sout{The latter} \mauro{Although both} agree reasonably well with $ {\cal T}_{12}$ and ${\cal T}_{34}$, how do we know for sure whether these are really the best match that we can get? This question is answered by plotting the concentration-dependent misfit function $\chi_{12}(n)$ and $\chi_{34}(n)$, which is shown in the inset of Fig.~\ref{fig1}(b). Both curves are obviously different but, reassuringly, display distinctive minima at the same concentration value of $N=50$, which coincides exactly with the actual impurity concentration contained in the parent configuration. This indicates that the inversion method introduced in \cite{shardul} can indeed account for the case of multiple electrodes.}

One question that may arise at this point is about the computational cost of calculating the misfit function since it contains an energy integral as well as an average involving $M$ configurations of disorder, not to mention the need to repeat the same procedure for a range of concentration values. Firstly, although expressed as an integral, the misfit function can be calculated as a discrete sum containing a reasonably small number of terms, usually $50$  or more, without compromising in accuracy. Regarding the CA procedure, we find that by taking $M$ $\approx$ $10^3$ configurations leads to results for the inverted concentrations with 
errors well below $5\%$, 
a remarkable achievement for a quantum inversion procedure \cite{shardul}.
%\shardul{This changes with the type of substrate as well as sampling region. For graphene and current device scheme this is correct} 
Finally, regarding the need to evaluate the misfit function for several concentration values, we adopt a machine-learning-based interpolating scheme \cite{Duarte_2021} that provides excellent resolution in $\chi_{1,2}(n)$ generated with only five distinct concentration values. 
%\section{Inversion Results with spatial resolution}
 
%\sout{Instead of using conductance readings between full-length electrode pairs that provide information about the overall impurity concentration, we may also interrogate the system with any pair of electrodes that do not necessarily span the entire device. Let us then consider the conductance ${\cal T}_{\widetilde{1},\widetilde{2}}(E)$ of the same device but this time current is being injected from the bottom left half-length electrode and extracted from its mirror-image electrode on the right, as shown in Fig.~\ref{fig2}(a).} 
Since it was possible to obtain the total number of impurities in the device with only a single reading (either ${\cal T}_{1,2}$ or ${\cal T}_{4,6}$), we assume that the use of additional input functions might enable us to spatially resolve how these impurities are distributed. With that in mind, we devise four partitions 
and assign the impurity numbers in each one of them as $N_1$ to $N_4$. It is important to stress that these are not physically built cells but simply a way in which we resolve the impurity number into smaller sections of the device. Vertical and horizontal dashed lines are drawn as a guide to the eyes in Fig.~\ref{fig1}(a) delineating the four distinct cells. 
% \sout{With the intention of resolving the impurity number in each of these cells, we label them $N_1$ to $N_4$, as indicated in Fig.~\ref{fig2}(a). For the sake of differentiation to the case of full-length electrodes, we use the symbol $\sim$ placed above the electrode-label integer to indicate that only a fraction of its length is active for current injection and extraction. In this case, the integers $\widetilde{j}$ represent electrodes with only half of their full length in contact with the conductor.} 
Similar steps can be taken in regard to the CA conductances in order to obtain the misfit functions defined in Eq.(\ref{misfit}) but this time the impurity numbers within the device may be broken into two separate parts: $N_{T}=N_1+N_2$ and $N_{B}=N_3+N_4$, {\it i.e.}, the top- and bottom-half values, respectively. This apparently adds one extra degree of freedom to the misfit function which should now read $\chi_{1,2}(N_{T},N_{B})$. Finding the minimum of $\chi_{1,2}(N_{T},N_{B})$ would in principle involve searching for minima of a two-variable function, but thanks to the previously obtained information about the total number of impurities in the device, we know that both variables are not independent but constrained to obey $N_{T} + N_{B} = N$. This is thus equivalent to a single-variable search. 
%------------------- FIG 2
\begin{figure}[!h]
    \centering
    \includegraphics[width=0.45\columnwidth]{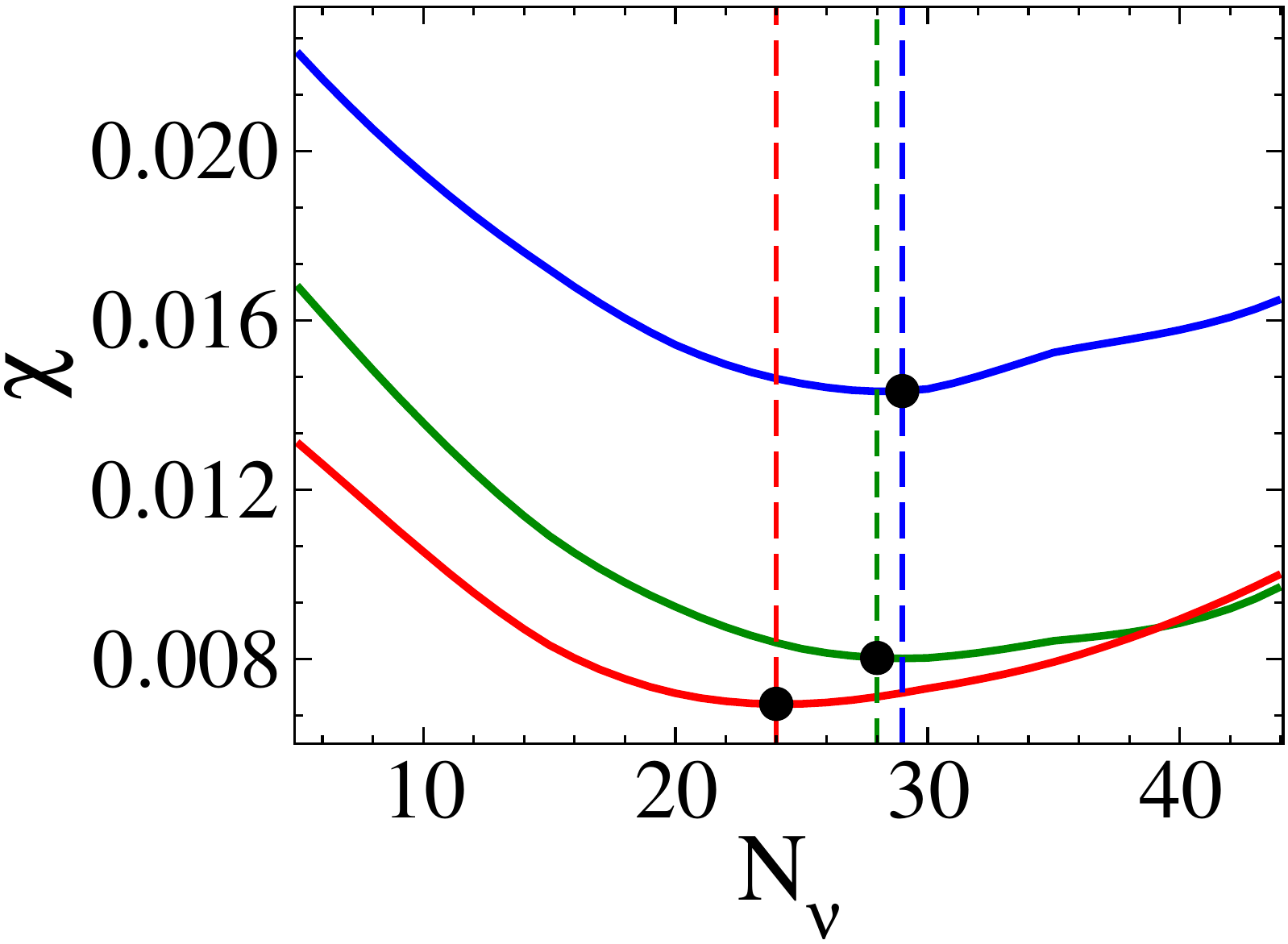}
    \caption{Misfit functions $\chi_{1,2}$ (green), $\chi_{5,6}$ (red), and $\chi_{1,3}$ (blue) plotted as a function of a generic variable $N_\nu$ representing the number of impurities at different parts of the device. $N_\nu = N_T$ for the green curve, $N_\nu = N_L$ for the red curve and $N_\nu = N_{D^\prime}$ for the blue curve. Vertical dashed lines indicate the distinctive minima seen in the misfit functions at $N_T=28$, $N_L=24$ and $N_{D^\prime}=29$ for the green, red and blue curves, respectively.}
    \label{fig2}
\end{figure}

In addition, the same can be done with horizontally placed electrodes. In this case ${\cal T}_{5,6}$ corresponds to the conductance between the top and bottom electrodes on the right of the device. Once again, two new variables are defined, namely, $N_{L}=N_1+N_3$ and $N_{R}=N_2+N_4$, which are the impurity numbers on the left and right halves of the device, respectively. The misfit function  $\chi_{4,6}(N_L,N_R)$ is another two-variable function that in practice depends only on one of them because they are also constrained to satisfy $N_L+N_R=N$. Fig.~\ref{fig2} depicts the  misfit functions  $\chi_{1,2}$ and $\chi_{5,6}$ plotted as a function of $N_T$ and $N_L$, respectively. Both curves display different minima located at values $N_T=28$ and $N_L=24$, respectively. Impurity numbers seen in Fig.~\ref{fig1}(a) may differ slightly from the occupation obtained from a single misfit-function minimum. Combined with the previously obtained result of $N=50$, in this case we may conclude that $N_B=22$ and $N_R=26$. Since $N_{T}$, $N_{B}$, $N_{L}$ and $N_{R}$ are not linearly independent quantities, knowing them
is not sufficient to uniquely identify the impurity numbers in each one of the four cells. For that an additional inversion involving the conductance between diagonally opposite electrodes is required, which is achieved with ${\cal T}_{1,3}$. A similar procedure leads to the corresponding misfit function $\chi_{1,3}(N_D,N_{D^\prime})$, 
where the variables $N_D=N_1+N_4$ and $N_{D^\prime}=N_2+N_3$ refer to the impurity numbers along the two diagonals. Obviously, they must also obey that $N_D+N_{D^\prime}=N$. Fig.~\ref{fig2} also shows $\chi_{1,3}$ plotted as a function of $N_{D^\prime}$, with a clear minimum at $N_{D^\prime}=29$ ($N_{D} = 21 $). 
The set of equations that cast these constraints is easily solved when expressed in matrix form, {\it i.e.}, 
\begin{equation}
\left(
\begin{array}{c}
N_1  \\
N_2  \\
N_3  \\
N_4 \\
\end{array}
\right) = \left(
\begin{array}{c c c c}
1 & 1 & 1 & 1 \\
1 & 1 & 0 & 0 \\
1 & 0 & 1 & 0 \\
1 & 0 & 0 & 1 \\
\end{array}
\right)^{-1}
 \left(
\begin{array}{c}
N  \\
N_T  \\
N_L  \\
N_D \\
\end{array}
\right) \,\,\,,
\label{sudoku-matrix}
\end{equation}
which leads to the impurity numbers on each of the four cells. 
In this case, $N_1=12.5$, $N_2=16.5$, $N_3=12.5$ and $N_4=9.5$. Interestingly, this is somewhat analogous to solving a Sudoku puzzle that starts from the conductance readings and finds the exact impurity numbers within each cell by imposing that they must add up to the specific values ($N_T$, $N_L$, $N_B$ and $N_D$) which are themselves determined through our inversion procedure.  

\begin{figure}[!h]
    \centering
    \includegraphics[width=0.8\columnwidth]{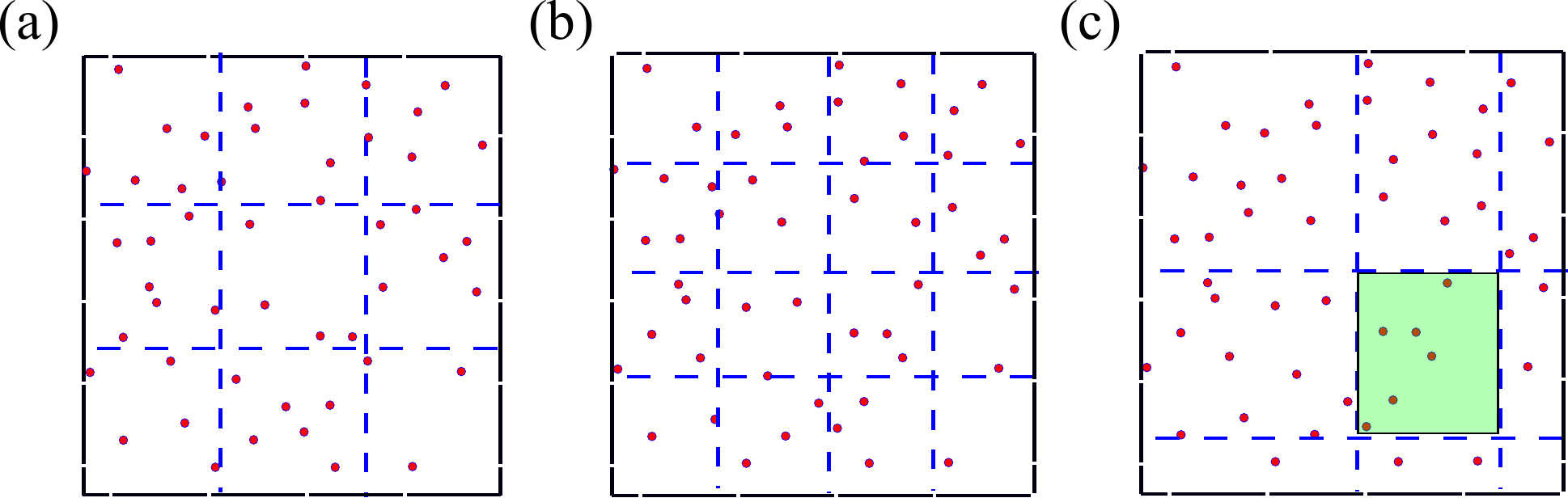}
    \caption{A schematic representation of the device resolved in (a) nine and (b) sixteen cells with the help of blue dashed lines. (c) Green region represents a cell of arbitrary size and location within the device.
    }
    \label{fig3}
\end{figure}

Three input functions (${\cal T}_{1,2}$, ${\cal T}_{5,6}$ and ${\cal T}_{1,3}$) were needed to find the impurity number in each of the four cells, {\it i.e.} two extra readings when compared to the inversion for finding the total number. This not only proves our earlier assumption that additional readings would enable us to spatially resolve the concentration but also suggests that an even better resolution is possible by interrogating the system further. It is important to stress that additional readings must be used in a hierarchical fashion, {\it i.e.}, they provide new information about the device always availing of results obtained by pre-existing readings, also offering constraints for a next higher resolution inversion procedure. In this way we are able to increase the resolution without the need for excessive new input functions but keeping them to a minimum. Following this path, we are able to go beyond the four-cell resolution.      
%\sout{Since by interrogating the disordered device with half-length electrodes we were able to spatially resolve the impurity concentration into four quarters, readings with even shorter electrodes are likely to improve the spatial resolution. In fact, the same Sudoku-style procedure can be carried out starting from conductance readings of ${\cal T}_{\widetilde{\widetilde{i}},\widetilde{\widetilde{j}}}$ where now the symbol $\approx$ above the integers in question are used to indicate that only one quarter of the $i$ and $j$ electrodes are active, as seen in Fig.~\ref{fig3}.} 
In this case a larger number of smaller cells is required, as seen in Fig.~\ref{fig3} where the same impurities are shown schematically without the underlying hexagonal atomic structure to avoid too congested a figure. Dashed lines in panels (a) and (b) are used to delineate nine and sixteen cells, respectively, each containing a certain fraction of the total number of impurities. 
%\sout{Double-sided arrows represent the quarter-length electrodes and the underlying hexagonal atomic structure has been omitted to avoid too congested a figure.} 
Our goal is to find the exact impurity number (concentration) in each one of these cells and the procedure is analogous to the case of four cells. The difference lies primarily in how the cells are combined to carry out the inversion. While the results of Fig.~\ref{fig2} were obtained by combining cells into rows and columns, there are very many ways of selecting how the individual cells can be clustered. This gives an enormous degree of flexibility on finding the actual number of impurities in each cell. Examples of a few different ways in which this is done can be found in the Supplemental Material (SM), but they all require five different readings, which once again involve an extra two input functions when compared to the four-cell case. For the sake of comparison, the boldface integers on the left and right parts of Table~\ref{table-9-16} indicate the impurity number $N_j$ in cell $j$ found through the Sudoku-style inversion in the case of nine and sixteen cells, respectively. Also shown in parenthesis is the real number of impurities contained in each cell. Note that the agreement is very good and that discrepancies rarely exceed a single unit. It is worth mentioning that natural uncertainties arise when impurities lie very close to the dashed lines separating the cells, since their scattering range may extend over more than one cell. This is particularly problematic as the cells are made smaller and the ratio between the cell perimeter and its area increases. Having mapped the concentration across the entire device in $3 \times 3$ or $4 \times 4$ cells, it is also possible to use this information to find the impurity number in a cell of arbitrary shape like the one shown in Fig.~\ref{fig3}(c) for instance. The size and location of that specific cell highlighted in green does not correspond to any of the cells seen in panels (a) and (b) and yet, the number of impurities occupying that specific cell can be found by carrying out a single CA calculation using the same input functions already utilised previously. As seen in the SM, this involves obtaining the minimum of a single-variable misfit function which, as previously mentioned, is not computationally intensive. 

% ----------------------- FIG. 3 ------------------- 

Up to this point, all results were based on the specific setup shown in Fig.\ref{fig1}(a) but equally successful inversions can be achieved with small variations, including electrodes of different shapes and sizes. The fact that our results and conclusions are not too sensitive to these variations indicates the robustness of our method. Furthermore, to demonstrate that there is great flexibility in how the device can be interrogated, we reproduce one of the cases shown previously and resolve the impurity concentration into four quadrants but this time we adopt a different setup for the electrodes, as shown in the SM. In this alternative setup electrodes are not of similar sizes, with one spanning the full-length of the device while the other is a quarter of its length. Reassuringly, this setup renders the same findings as the ones obtained with the dashed-line cells defined in Fig.~\ref{fig1}(a). That degree of flexibility suggests that a wide range of different possibilities exists to probe the local conductances of disordered 2D materials in order to map how their impurities and defects are spatially distributed.

\begin{table}
    \centering
    \begin{tabular}{ |P{1.05cm} |P{1.05cm}| P{1.05cm}|}
    \hline
    {\bf 7} (6) &  {\bf 8} (7) & {\bf 5} (6) \\[4.2pt]
    \hline
    {\bf 8} (7) & {\bf 5} (4) & {\bf 7} (6)  \\[4.2pt]
    \hline
    {\bf 4} (5)   & {\bf 5} (6) & {\bf 3} (3) \\[4.2pt]
    \hline
   \end{tabular}
%\end{table}
%\begin{table}
    \begin{tabular}{ | P{1.05cm} |P{1.05cm} | P{1.05cm} |P{1.05cm} |}
    \hline
    {\bf 0} (2) &  {\bf 2} (3) & {\bf 5} (4) & {\bf 3} (4)\\
    \hline
    {\bf 7} (5) & {\bf 4} (3) & {\bf 1} (2) & {\bf 3} (3) \\
    \hline
    {\bf 5} (5)  & {\bf 2} (3) & {\bf 6} (4) & {\bf 2} (2) \\
    \hline
        {\bf 3} (2)   & {\bf 2} (3) & {\bf 3} (3) & {\bf 2} (2) \\
    \hline
   \end{tabular}
   \caption{Boldface integers indicate the number of impurities obtained from the Sudoku-style inversion. The accompanying integers in parenthesis indicate the real number as per Fig.~\ref{fig3}. Left: 3 x 3; Right: 4 x 4}
   \label{table-9-16}
\end{table}
%Following our notation, Fig.~\ref{fig1}b shows the conductance readings  ${\cal T}_{1,\widetilde{\widetilde{2}}}$ and ${\cal T}_{3,\widetilde{\widetilde{4}}}$ together with the corresponding misfit functions $\chi_{1,\widetilde{\widetilde{2}}}$ and $\chi_{3,\widetilde{\widetilde{4}}}$ in the inset. While the conductance readings and the corresponding misfit functions differ from the cases seen in Fig. \ref{fig2}b, the minimizing values of the misfit function are exactly the same, leading to the same four occupation values found previously even though an entirely distinct electrode setup was used.    

We test the inversion accuracy by defining the error as $\alpha = {1 \over J} \sum_j^J \vert N_j^{\rm min} - N_j^{P} \vert /N_j^{P}$, where $N_j^{P}$ is the number of impurities in cell $j$ of the parent (input) configuration and $N_j^{\rm min}$ is the corresponding value that minimizes the misfit function. $\alpha$ consists of an average over the total number of cells $J$, so that large values of $J$ correspond to higher spatial resolution in the impurity distribution. Having considered the cases of $J=1$, $J=4$, $J=9$ and $J=16$, we may conclude that the inversion accuracy decreases as we attempt to increase the resolution, as seen in Table \ref{tab:my_label}. These values are obtained out of repeated inversions from $100$
different parent configurations in order to achieve statistical significance. The higher the number of cells the higher the resolution but, unfortunately, the lower the accuracy (roughly a $1/\sqrt{N}$ scaling, where $N$ is the average number of defects in a cell).
Given the diverse ways in which a device can be probed and interrogated, it is possible that by increasing the number of readings and/or by selecting more appropriate electrode setups, the accuracy can be further improved. A brief account of the method scalability appears in the SM. Remarkably, moderate resolution can be achieved in mapping how impurities are distributed with this inversion methodology. 

\section{Conclusion and outlook}

The Sudoku-style inversion tool presented here has been based entirely on multi-terminal conductance measurements
serving as input functions. However, the basic non-spatially-resolving inversion has been shown to work with other input signals\cite{shardul, Duarte_2021}. In fact, other quantities that can be written in terms of two-point correlation functions are likely candidates to display similar characteristics in the presence of disorder and, therefore, may serve as potential input functions from which spatial mapping of impurity concentration becomes possible. This may pave the way to using thermal conductivities, spin susceptibilities, to name but a few, as sources capable of providing spatial information about a disordered device when only moderate resolution is required. 

In summary, we have shown how a recently proposed inversion methodology capable of identifying the global 
number of impurities from seemingly noisy two-terminal conductance signals of a 2D quantum device can be extended to determine how such impurities are spatially distributed. By generalising the method to a multi-terminal framework and 
mapping the device into a grid-like structure, the inversion specifies the total impurity number in separate rows and columns. These are constraints that must be satisfied in order to find the exact concentration in each one of the individual cells, in a way that resembles a Sudoku puzzle. The impurity distribution can be resolved into smaller sections of the device, depending on how the local conductance is being probed. %\sout{and on the dimensions of the probing electrodes.}
Furthermore, spatial resolution can certainly be improved by increasing the number of conductance readings. We envisage this inversion methodology being implemented beyond conductance measurements which will provide a framework for visualising impurities from simple measurements of different physical properties. 

\begin{table}
    \centering
    \begin{tabular}{ | P{2.5cm} |P{2.5cm} | }
    \hline
    $\#$ of cells  $J$ &  Error ($\alpha$) \\
    \hline
    1  & 0.037  \\
    \hline
    4   & 0.12  \\
    \hline
    9 & 0.17\\
    \hline
    16   & 0.21  \\
   \hline 
   \end{tabular}
   \caption{Inversion method accuracy $\alpha$ for different number of cells $J$. }
   \label{tab:my_label}
\end{table}

\section{Acknowledgments}
This publication has emanated from research supported in part by a research grant from Science Foundation Ireland (SFI) under Grant Number SFI/12/RC/2278-P2. C.L acknowledges financial support from CNPq under Grant \#313059/2020-9 and from FAPERJ under Grant \#E-26/202.882/2018. 

%\begin{figure}
%    \includegraphics[width=\columnwidth]{figures/4termi_sudo_q_1.svg.pdf}
%    \caption{L= 100; 1.5\% impurities randomly distributed inside the flake}
%\end{figure}
%\begin{figure}
%    \centering
%    \includegraphics[width=\columnwidth]{figures/sudo_q_result_1.pdf}
%    \caption{SOLVED SUDO Q}
%\end{figure}

%--------------------------------
\bibliography{mybib,dftbib}

\begin{thebibliography}{10}
\expandafter\ifx\csname url\endcsname\relax
  \def\url#1{\texttt{#1}}\fi
\expandafter\ifx\csname urlprefix\endcsname\relax\def\urlprefix{URL }\fi
\expandafter\ifx\csname href\endcsname\relax
  \def\href#1#2{#2} \def\path#1{#1}\fi

\bibitem{medical}
M.~Bertero, M.~Piana, Inverse problems in biomedical imaging: modeling and
  methods of solution, Springer Milan, Milano, 2006, pp. 1--33.

\bibitem{fwi}
J.~Virieux, A.~Asnaashari, R.~Brossier, L.~Métivier, A.~Ribodetti, W.~Zhou,
  \href{https://library.seg.org/doi/abs/10.1190/1.9781560803027.entry6}{{A}n
  introduction to full waveform inversion}, 2017, pp. R1--1--R1--40.
\newblock \href {http://dx.doi.org/10.1190/1.9781560803027.entry6}
  {\path{doi:10.1190/1.9781560803027.entry6}}.
\newline\urlprefix\url{https://library.seg.org/doi/abs/10.1190/1.9781560803027.entry6}

\bibitem{tromp2008spectral}
J.~Tromp, D.~Komatitsch, Q.~Liu,
  \href{http://global-sci.org/intro/article_detail/cicp/7840.html}{Spectral-element
  and adjoint methods in seismology}, Communications in Computational Physics
  3~(1) (2008) 1--32.
\newline\urlprefix\url{http://global-sci.org/intro/article_detail/cicp/7840.html}

\bibitem{sonar}
E.~T.~F. Dias, H.~Vieira~Neto, A novel approach to environment mapping using
  sonar sensors and inverse problems, in: C.~Dixon, K.~Tuyls (Eds.), Towards
  Autonomous Robotic Systems, Springer International Publishing, Cham, 2015,
  pp. 100--111.

\bibitem{Lassas2008}
M.~Lassas, L.~P{\"a}iv{\"a}rinta, E.~Saksman,
  \href{https://doi.org/10.1007/s00220-008-0416-6}{Inverse scattering problem
  for a two dimensional random potential}, Commun. Math. Phys. 279~(3) (2008)
  669--703.
\newblock \href {http://dx.doi.org/10.1007/s00220-008-0416-6}
  {\path{doi:10.1007/s00220-008-0416-6}}.
\newline\urlprefix\url{https://doi.org/10.1007/s00220-008-0416-6}

\bibitem{PhysRevLett.111.090403}
T.~M. Hoang, C.~S. Gerving, B.~J. Land, M.~Anquez, C.~D. Hamley, M.~S. Chapman,
  \href{https://link.aps.org/doi/10.1103/PhysRevLett.111.090403}{Dynamic
  stabilization of a quantum many-body spin system}, Phys. Rev. Lett. 111
  (2013) 090403.
\newblock \href {http://dx.doi.org/10.1103/PhysRevLett.111.090403}
  {\path{doi:10.1103/PhysRevLett.111.090403}}.
\newline\urlprefix\url{https://link.aps.org/doi/10.1103/PhysRevLett.111.090403}

\bibitem{PhysRevX.8.031029}
E.~Chertkov, B.~K. Clark,
  \href{https://link.aps.org/doi/10.1103/PhysRevX.8.031029}{Computational
  inverse method for constructing spaces of quantum models from wave
  functions}, Phys. Rev. X 8 (2018) 031029.
\newblock \href {http://dx.doi.org/10.1103/PhysRevX.8.031029}
  {\path{doi:10.1103/PhysRevX.8.031029}}.
\newline\urlprefix\url{https://link.aps.org/doi/10.1103/PhysRevX.8.031029}

\bibitem{Lai_2017}
R.-Y. Lai, R.~Shankar, D.~Spirn, G.~Uhlmann,
  \href{https://doi.org/10.1088/1361-6420/aa8e81}{An inverse problem from
  condensed matter physics}, Inverse Problems 33~(11) (2017) 115011.
\newblock \href {http://dx.doi.org/10.1088/1361-6420/aa8e81}
  {\path{doi:10.1088/1361-6420/aa8e81}}.
\newline\urlprefix\url{https://doi.org/10.1088/1361-6420/aa8e81}

\bibitem{tsymbal}
E.~Tsymbal, P.~Dowben,
  \href{https://www.frontiersin.org/article/10.3389/fphy.2013.00032}{Grand
  challenges in condensed matter physics: from knowledge to innovation},
  Frontiers in Physics 1 (2013) 32.
\newblock \href {http://dx.doi.org/10.3389/fphy.2013.00032}
  {\path{doi:10.3389/fphy.2013.00032}}.
\newline\urlprefix\url{https://www.frontiersin.org/article/10.3389/fphy.2013.00032}

\bibitem{jasper}
J.~van~der Gucht,
  \href{https://www.frontiersin.org/article/10.3389/fphy.2018.00087}{Grand
  challenges in soft matter physics}, Frontiers in Physics 6 (2018) 87.
\newblock \href {http://dx.doi.org/10.3389/fphy.2018.00087}
  {\path{doi:10.3389/fphy.2018.00087}}.
\newline\urlprefix\url{https://www.frontiersin.org/article/10.3389/fphy.2018.00087}

\bibitem{gianluca}
G.~Bertaina, D.~E. Galli, E.~Vitali,
  \href{https://doi.org/10.1080/23746149.2017.1288585}{Statistical and
  computational intelligence approach to analytic continuation in {Quantum
  Monte Carlo}}, Advances in Physics: X 2~(2) (2017) 302--323.
\newblock \href {http://dx.doi.org/10.1080/23746149.2017.1288585}
  {\path{doi:10.1080/23746149.2017.1288585}}.
\newline\urlprefix\url{https://doi.org/10.1080/23746149.2017.1288585}

\bibitem{Franceschetti1999}
A.~Franceschetti, A.~Zunger, \href{https://doi.org/10.1038/46995}{The inverse
  band-structure problem of finding an atomic configuration with given
  electronic properties}, Nature 402~(6757) (1999) 60--63.
\newblock \href {http://dx.doi.org/10.1038/46995} {\path{doi:10.1038/46995}}.
\newline\urlprefix\url{https://doi.org/10.1038/46995}

\bibitem{liping}
L.~Yu, R.~S. Kokenyesi, D.~A. Keszler, A.~Zunger,
  \href{https://onlinelibrary.wiley.com/doi/abs/10.1002/aenm.201200538}{Inverse
  design of high absorption thin-film photovoltaic materials}, Adv. Energy
  Mater. 3~(1) (2013) 43--48.
\newblock \href {http://dx.doi.org/10.1002/aenm.201200538}
  {\path{doi:10.1002/aenm.201200538}}.
\newline\urlprefix\url{https://onlinelibrary.wiley.com/doi/abs/10.1002/aenm.201200538}

\bibitem{plasma}
M.~F. Kasim, T.~Galligan, J.~T. Mugglestone, G.~Gregori, S.~M. Vinko, Inverse
  problem instabilities in large-scale modelling of matter in extreme
  conditions, Phys. Plasmas 26 (2019) 112706.
\newblock \href {http://dx.doi.org/10.1063/1.5125979}
  {\path{doi:10.1063/1.5125979}}.

\bibitem{jensen}
D.~S. Jensen, A.~Wasserman,
  \href{https://onlinelibrary.wiley.com/doi/abs/10.1002/qua.25425}{Numerical
  methods for the inverse problem of density functional theory}, Int. J.
  Quantum Chem. 118~(1) (2018) e25425.
\newblock \href {http://dx.doi.org/10.1002/qua.25425}
  {\path{doi:10.1002/qua.25425}}.
\newline\urlprefix\url{https://onlinelibrary.wiley.com/doi/abs/10.1002/qua.25425}

\bibitem{yazyev}
R.~Fournier, L.~Wang, O.~V. Yazyev, Q.~Wu,
  \href{https://link.aps.org/doi/10.1103/PhysRevLett.124.056401}{Artificial
  neural network approach to the analytic continuation problem}, Phys. Rev.
  Lett. 124 (2020) 056401.
\newblock \href {http://dx.doi.org/10.1103/PhysRevLett.124.056401}
  {\path{doi:10.1103/PhysRevLett.124.056401}}.
\newline\urlprefix\url{https://link.aps.org/doi/10.1103/PhysRevLett.124.056401}

\bibitem{Zhang2013}
L.~Zhang, J.-W. Luo, A.~Saraiva, B.~Koiller, A.~Zunger,
  \href{https://doi.org/10.1038/ncomms3396}{Genetic design of enhanced valley
  splitting towards a spin qubit in silicon}, Nat. Commun. 4~(1) (2013) 2396.
\newblock \href {http://dx.doi.org/10.1038/ncomms3396}
  {\path{doi:10.1038/ncomms3396}}.
\newline\urlprefix\url{https://doi.org/10.1038/ncomms3396}

\bibitem{Vargas2019}
R.~A. Vargas-Hern{\'{a}}ndez, Y.~Guan, D.~H. Zhang, R.~V. Krems,
  \href{https://doi.org/10.1088%2F1367-2630%2Fab0099}{Bayesian optimization for
  the inverse scattering problem in quantum reaction dynamics}, New J. Phys.
  21~(2) (2019) 022001.
\newblock \href {http://dx.doi.org/10.1088/1367-2630/ab0099}
  {\path{doi:10.1088/1367-2630/ab0099}}.
\newline\urlprefix\url{https://doi.org/10.1088%2F1367-2630%2Fab0099}

\bibitem{kyriienko}
O.~Kyriienko, Quantum inverse iteration algorithm for programmable quantum
  simulators, npj Quantum Information 6.
\newblock \href {http://dx.doi.org/10.1038/s41534-019-0239-7}
  {\path{doi:10.1038/s41534-019-0239-7}}.

\bibitem{anatole}
A.~Lopez-Bezanilla, O.~A. von Lilienfeld,
  \href{https://link.aps.org/doi/10.1103/PhysRevB.89.235411}{Modeling
  electronic quantum transport with machine learning}, Phys. Rev. B 89 (2014)
  235411.
\newblock \href {http://dx.doi.org/10.1103/PhysRevB.89.235411}
  {\path{doi:10.1103/PhysRevB.89.235411}}.
\newline\urlprefix\url{https://link.aps.org/doi/10.1103/PhysRevB.89.235411}

\bibitem{fazli}
K.~Hansen, G.~Montavon, F.~Biegler, S.~Fazli, M.~Rupp, M.~Scheffler, O.~A. von
  Lilienfeld, A.~Tkatchenko, K.-R. Müller,
  \href{https://doi.org/10.1021/ct400195d}{Assessment and validation of machine
  learning methods for predicting molecular atomization energies}, J. Chem.
  Theory Comput. 9~(8) (2013) 3404--3419.
\newblock \href {http://dx.doi.org/10.1021/ct400195d}
  {\path{doi:10.1021/ct400195d}}.
\newline\urlprefix\url{https://doi.org/10.1021/ct400195d}

\bibitem{burak}
B.~Himmetoglu, \href{https://doi.org/10.1063/1.4964093}{Tree based machine
  learning framework for predicting ground state energies of molecules}, J.
  Chem. Phys. 145~(13) (2016) 134101.
\newblock \href {http://dx.doi.org/10.1063/1.4964093}
  {\path{doi:10.1063/1.4964093}}.
\newline\urlprefix\url{https://doi.org/10.1063/1.4964093}

\bibitem{collins}
H.~Li, C.~Collins, M.~Tanha, G.~J. Gordon, D.~J. Yaron, A density functional
  tight binding layer for deep learning of chemical {H}amiltonians, J. Chem.
  Theory Comput. 14 (2018) 5764.
\newblock \href {http://dx.doi.org/10.1021/acs.jctc.8b00873}
  {\path{doi:10.1021/acs.jctc.8b00873}}.

\bibitem{Xia2018}
R.~Xia, S.~Kais, \href{https://doi.org/10.1038/s41467-018-06598-z}{Quantum
  machine learning for electronic structure calculations}, Nat. Commun. 9~(1)
  (2018) 4195.
\newblock \href {http://dx.doi.org/10.1038/s41467-018-06598-z}
  {\path{doi:10.1038/s41467-018-06598-z}}.
\newline\urlprefix\url{https://doi.org/10.1038/s41467-018-06598-z}

\bibitem{dral}
P.~O. Dral, \href{https://doi.org/10.1021/acs.jpclett.9b03664}{Quantum
  chemistry in the age of machine learning}, J. Phys. Chem. Lett. 11~(6) (2020)
  2336--2347.
\newblock \href {http://dx.doi.org/10.1021/acs.jpclett.9b03664}
  {\path{doi:10.1021/acs.jpclett.9b03664}}.
\newline\urlprefix\url{https://doi.org/10.1021/acs.jpclett.9b03664}

\bibitem{melko}
J.~Carrasquilla, R.~G. Melko, \href{https://doi.org/10.1038/nphys4035}{Machine
  learning phases of matter}, Nat. Phys. 13~(5) (2017) 431--434.
\newblock \href {http://dx.doi.org/10.1038/nphys4035}
  {\path{doi:10.1038/nphys4035}}.
\newline\urlprefix\url{https://doi.org/10.1038/nphys4035}

\bibitem{Fazzio2019}
G.~R. {Schleder}, A.~C.~M. {Padilha}, C.~M. {Acosta}, M.~{Costa}, A.~{Fazzio},
  {From DFT to machine learning: recent approaches to materials science-a
  review}, Journal of Physics: Materials 2~(3) (2019) 032001.
\newblock \href {http://dx.doi.org/10.1088/2515-7639/ab084b}
  {\path{doi:10.1088/2515-7639/ab084b}}.

\bibitem{Ziletti2018}
A.~Ziletti, D.~Kumar, M.~Scheffler, L.~M. Ghiringhelli,
  \href{https://doi.org/10.1038/s41467-018-05169-6}{Insightful classification
  of crystal structures using deep learning}, Nat. Commun. 9~(1) (2018) 2775.
\newblock \href {http://dx.doi.org/10.1038/s41467-018-05169-6}
  {\path{doi:10.1038/s41467-018-05169-6}}.
\newline\urlprefix\url{https://doi.org/10.1038/s41467-018-05169-6}

\bibitem{rajan}
R.~Potyrailo, K.~Rajan, K.~Stoewe, I.~Takeuchi, B.~Chisholm, H.~Lam,
  \href{https://doi.org/10.1021/co200007w}{Combinatorial and high-throughput
  screening of materials libraries: Review of state of the art}, ACS Comb. Sci.
  13~(6) (2011) 579--633.
\newblock \href {http://dx.doi.org/10.1021/co200007w}
  {\path{doi:10.1021/co200007w}}.
\newline\urlprefix\url{https://doi.org/10.1021/co200007w}

\bibitem{suram}
S.~K. Suram, P.~F. Newhouse, L.~Zhou, D.~G. Van~Campen, A.~Mehta, J.~M.
  Gregoire, \href{https://doi.org/10.1021/acscombsci.6b00054}{High throughput
  light absorber discovery, part 2: Establishing structure–band gap energy
  relationships}, ACS Comb. Sci. 18~(11) (2016) 682--688.
\newblock \href {http://dx.doi.org/10.1021/acscombsci.6b00054}
  {\path{doi:10.1021/acscombsci.6b00054}}.
\newline\urlprefix\url{https://doi.org/10.1021/acscombsci.6b00054}

\bibitem{Koinuma2004}
H.~Koinuma, I.~Takeuchi, \href{https://doi.org/10.1038/nmat1157}{Combinatorial
  solid-state chemistry of inorganic materials}, Nat. Mater. 3~(7) (2004)
  429--438.
\newblock \href {http://dx.doi.org/10.1038/nmat1157}
  {\path{doi:10.1038/nmat1157}}.
\newline\urlprefix\url{https://doi.org/10.1038/nmat1157}

\bibitem{choi}
M.~L. Green, C.~L. Choi, J.~R. Hattrick-Simpers, A.~M. Joshi, I.~Takeuchi,
  S.~C. Barron, E.~Campo, T.~Chiang, S.~Empedocles, J.~M. Gregoire, A.~G.
  Kusne, J.~Martin, A.~Mehta, K.~Persson, Z.~Trautt, J.~Van~Duren,
  A.~Zakutayev, \href{https://doi.org/10.1063/1.4977487}{Fulfilling the promise
  of the materials genome initiative with high-throughput experimental
  methodologies}, Appl. Phys. Rev. 4~(1) (2017) 011105.
\newblock \href {http://dx.doi.org/10.1063/1.4977487}
  {\path{doi:10.1063/1.4977487}}.
\newline\urlprefix\url{https://doi.org/10.1063/1.4977487}

\bibitem{nardelli}
S.~Curtarolo, G.~L.~W. Hart, M.~B. Nardelli, N.~Mingo, S.~Sanvito, O.~Levy,
  \href{https://doi.org/10.1038/nmat3568}{The high-throughput highway to
  computational materials design}, Nat. Mater. 12~(3) (2013) 191--201.
\newblock \href {http://dx.doi.org/10.1038/nmat3568}
  {\path{doi:10.1038/nmat3568}}.
\newline\urlprefix\url{https://doi.org/10.1038/nmat3568}

\bibitem{PhysRevLett.108.068701}
L.~Yu, A.~Zunger,
  \href{https://link.aps.org/doi/10.1103/PhysRevLett.108.068701}{Identification
  of potential photovoltaic absorbers based on first-principles spectroscopic
  screening of materials}, Phys. Rev. Lett. 108 (2012) 068701.
\newblock \href {http://dx.doi.org/10.1103/PhysRevLett.108.068701}
  {\path{doi:10.1103/PhysRevLett.108.068701}}.
\newline\urlprefix\url{https://link.aps.org/doi/10.1103/PhysRevLett.108.068701}

\bibitem{Fischer2006}
C.~C. Fischer, K.~J. Tibbetts, D.~Morgan, G.~Ceder,
  \href{https://doi.org/10.1038/nmat1691}{Predicting crystal structure by
  merging data mining with quantum mechanics}, Nat. Mater. 5~(8) (2006)
  641--646.
\newblock \href {http://dx.doi.org/10.1038/nmat1691}
  {\path{doi:10.1038/nmat1691}}.
\newline\urlprefix\url{https://doi.org/10.1038/nmat1691}

\bibitem{Gautier2015}
R.~Gautier, X.~Zhang, L.~Hu, L.~Yu, Y.~Lin, T.~O.~L. Sunde, D.~Chon, K.~R.
  Poeppelmeier, A.~Zunger, \href{https://doi.org/10.1038/nchem.2207}{Prediction
  and accelerated laboratory discovery of previously unknown 18-electron abx
  compounds}, Nat. Chem. 7~(4) (2015) 308--316.
\newblock \href {http://dx.doi.org/10.1038/nchem.2207}
  {\path{doi:10.1038/nchem.2207}}.
\newline\urlprefix\url{https://doi.org/10.1038/nchem.2207}

\bibitem{Schmidt2019}
J.~Schmidt, M.~R.~G. Marques, S.~Botti, M.~A.~L. Marques,
  \href{https://doi.org/10.1038/s41524-019-0221-0}{Recent advances and
  applications of machine learning in solid-state materials science}, npj
  Computational Materials 5~(1) (2019) 83.
\newblock \href {http://dx.doi.org/10.1038/s41524-019-0221-0}
  {\path{doi:10.1038/s41524-019-0221-0}}.
\newline\urlprefix\url{https://doi.org/10.1038/s41524-019-0221-0}

\bibitem{Lewenkopf2010}
E.~R. Mucciolo, C.~H. Lewenkopf,
  \href{https://doi.org/10.1088/0953-8984/22/27/273201}{Disorder and electronic
  transport in graphene}, J. Phys.: Condens. Matter 22~(27) (2010) 273201.
\newblock \href {http://dx.doi.org/10.1088/0953-8984/22/27/273201}
  {\path{doi:10.1088/0953-8984/22/27/273201}}.
\newline\urlprefix\url{https://doi.org/10.1088/0953-8984/22/27/273201}

\bibitem{DasSarma2011}
S.~Das~Sarma, S.~Adam, E.~H. Hwang, E.~Rossi,
  \href{https://link.aps.org/doi/10.1103/RevModPhys.83.407}{Electronic
  transport in two-dimensional graphene}, Rev. Mod. Phys. 83 (2011) 407--470.
\newblock \href {http://dx.doi.org/10.1103/RevModPhys.83.407}
  {\path{doi:10.1103/RevModPhys.83.407}}.
\newline\urlprefix\url{https://link.aps.org/doi/10.1103/RevModPhys.83.407}

\bibitem{shardul}
S.~Mukim, F.~P. Amorim, A.~R. Rocha, R.~B. Muniz, C.~Lewenkopf, M.~S. Ferreira,
  \href{https://link.aps.org/doi/10.1103/PhysRevB.102.075409}{Disorder
  information from conductance: A quantum inverse problem}, Phys. Rev. B 102
  (2020) 075409.
\newblock \href {http://dx.doi.org/10.1103/PhysRevB.102.075409}
  {\path{doi:10.1103/PhysRevB.102.075409}}.
\newline\urlprefix\url{https://link.aps.org/doi/10.1103/PhysRevB.102.075409}

\bibitem{Ihn2010}
T.~Ihn, {Semiconductor Nanostructures}, Oxford University Press, Oxford, 2010.
\newblock \href {http://dx.doi.org/10.1093/acprof:oso/9780199534425.001.0001}
  {\path{doi:10.1093/acprof:oso/9780199534425.001.0001}}.

\bibitem{Ghahari2011}
F.~Ghahari, Y.~Zhao, P.~Cadden-Zimansky, K.~Bolotin, P.~Kim,
  \href{https://link.aps.org/doi/10.1103/PhysRevLett.106.046801}{Measurement of
  the $\ensuremath{\nu}=1/3$ fractional quantum hall energy gap in suspended
  graphene}, Phys. Rev. Lett. 106 (2011) 046801.
\newblock \href {http://dx.doi.org/10.1103/PhysRevLett.106.046801}
  {\path{doi:10.1103/PhysRevLett.106.046801}}.
\newline\urlprefix\url{https://link.aps.org/doi/10.1103/PhysRevLett.106.046801}

\bibitem{Perkins2013}
E.~Perkins, L.~Barreto, J.~Wells, P.~Hofmann,
  \href{https://doi.org/10.1063/1.4793376}{Surface-sensitive conductivity
  measurement using a micro multi-point probe approach}, Rev. Sci. Instrum.
  84~(3) (2013) 033901.
\newblock \href {http://dx.doi.org/10.1063/1.4793376}
  {\path{doi:10.1063/1.4793376}}.
\newline\urlprefix\url{https://doi.org/10.1063/1.4793376}

\bibitem{doi:10.1021/acsnano.6b05288}
H.~Lee, D.~Cho, S.~Shekhar, J.~Kim, J.~Park, B.~H. Hong, S.~Hong,
  \href{https://doi.org/10.1021/acsnano.6b05288}{Nanoscale direct mapping of
  noise source activities on graphene domains}, ACS Nano 10~(11) (2016)
  10135--10142, pMID: 27934081.
\newblock \href {http://arxiv.org/abs/https://doi.org/10.1021/acsnano.6b05288}
  {\path{arXiv:https://doi.org/10.1021/acsnano.6b05288}}, \href
  {http://dx.doi.org/10.1021/acsnano.6b05288}
  {\path{doi:10.1021/acsnano.6b05288}}.
\newline\urlprefix\url{https://doi.org/10.1021/acsnano.6b05288}

\bibitem{Cui2015}
X.~Cui, G.-H. Lee, Y.~D. Kim, G.~Arefe, P.~Y. Huang, C.-H. Lee, D.~A. Chenet,
  X.~Zhang, L.~Wang, F.~Ye, F.~Pizzocchero, B.~S. Jessen, K.~Watanabe,
  T.~Taniguchi, D.~A. Muller, T.~Low, P.~Kim, J.~Hone,
  \href{https://doi.org/10.1038/nnano.2015.70}{Multi-terminal transport
  measurements of {MoS}$_2$ using a van der waals heterostructure device
  platform}, Nature Nanotechnology 10~(6) (2015) 534--540.
\newblock \href {http://dx.doi.org/10.1038/nnano.2015.70}
  {\path{doi:10.1038/nnano.2015.70}}.
\newline\urlprefix\url{https://doi.org/10.1038/nnano.2015.70}

\bibitem{SHIN201959}
N.~Shin, J.~Kim, S.~Shekhar, M.~Yang, S.~Hong,
  \href{https://www.sciencedirect.com/science/article/pii/S0008622318308431}{Nanoscale
  reduction of resistivity and charge trap activities induced by carbon
  nanotubes embedded in metal thin films}, Carbon 141 (2019) 59--66.
\newblock \href
  {http://dx.doi.org/https://doi.org/10.1016/j.carbon.2018.09.029}
  {\path{doi:https://doi.org/10.1016/j.carbon.2018.09.029}}.
\newline\urlprefix\url{https://www.sciencedirect.com/science/article/pii/S0008622318308431}

\bibitem{Buttiker1986}
M.~B\"uttiker,
  \href{https://link.aps.org/doi/10.1103/PhysRevLett.57.1761}{Four-terminal
  phase-coherent conductance}, Phys. Rev. Lett. 57 (1986) 1761--1764.
\newblock \href {http://dx.doi.org/10.1103/PhysRevLett.57.1761}
  {\path{doi:10.1103/PhysRevLett.57.1761}}.
\newline\urlprefix\url{https://link.aps.org/doi/10.1103/PhysRevLett.57.1761}

\bibitem{Jura2007}
M.~Jura, M.~Topinka, L.~Urban, A.~Yazdani, H.~Shtrikman, L.~N. Pfeiffer, K.~W.
  West, D.~Goldhaber-Gordon, Unexpected features of branched flow through
  high-mobility two-dimensional electron gases, Nat. Phys. 3 (2007) 841–845.
\newblock \href {http://dx.doi.org/10.1038/nphys756}
  {\path{doi:10.1038/nphys756}}.

\bibitem{Braem2018}
B.~A. Braem, F.~M.~D. Pellegrino, A.~Principi, M.~R\"o\"osli, C.~Gold,
  S.~Hennel, J.~V. Koski, M.~Berl, W.~Dietsche, W.~Wegscheider, M.~Polini,
  T.~Ihn, K.~Ensslin,
  \href{https://link.aps.org/doi/10.1103/PhysRevB.98.241304}{Scanning gate
  microscopy in a viscous electron fluid}, Phys. Rev. B 98 (2018) 241304.
\newblock \href {http://dx.doi.org/10.1103/PhysRevB.98.241304}
  {\path{doi:10.1103/PhysRevB.98.241304}}.
\newline\urlprefix\url{https://link.aps.org/doi/10.1103/PhysRevB.98.241304}

\bibitem{Brun2019}
B.~Brun, N.~Moreau, S.~Somanchi, V.-H. Nguyen, K.~Watanabe, T.~Taniguchi, J.-C.
  Charlier, C.~Stampfer, B.~Hackens,
  \href{https://link.aps.org/doi/10.1103/PhysRevB.100.041401}{Imaging {D}irac
  fermions flow through a circular {V}eselago lens}, Phys. Rev. B 100 (2019)
  041401.
\newblock \href {http://dx.doi.org/10.1103/PhysRevB.100.041401}
  {\path{doi:10.1103/PhysRevB.100.041401}}.
\newline\urlprefix\url{https://link.aps.org/doi/10.1103/PhysRevB.100.041401}

\bibitem{Brun2020}
B.~Brun, N.~Moreau, S.~Somanchi, V.-H. Nguyen,
  A.~Mre{\'{n}}ca-Kolasi{\'{n}}ska, K.~Watanabe, T.~Taniguchi, J.-C. Charlier,
  C.~Stampfer, B.~Hackens,
  \href{https://doi.org/10.1088/2053-1583/ab734e}{Optimizing {D}irac fermions
  quasi-confinement by potential smoothness engineering}, 2D Materials 7~(2)
  (2020) 025037.
\newblock \href {http://dx.doi.org/10.1088/2053-1583/ab734e}
  {\path{doi:10.1088/2053-1583/ab734e}}.
\newline\urlprefix\url{https://doi.org/10.1088/2053-1583/ab734e}

\bibitem{Ping2013}
A.-P. Li, K.~W. Clark, X.-G. Zhang, A.~P. Baddorf,
  \href{https://onlinelibrary.wiley.com/doi/abs/10.1002/adfm.201203423}{Electron
  transport at the nanometer-scale spatially revealed by four-probe scanning
  tunneling microscopy}, Advanced Functional Materials 23~(20) (2013)
  2509--2524.
\newblock \href {http://dx.doi.org/https://doi.org/10.1002/adfm.201203423}
  {\path{doi:https://doi.org/10.1002/adfm.201203423}}.
\newline\urlprefix\url{https://onlinelibrary.wiley.com/doi/abs/10.1002/adfm.201203423}

\bibitem{Baringhaus2014}
J.~Baringhaus, M.~Ruan, F.~Edler, A.~Tejeda, M.~Sicot, A.~Taleb-Ibrahimi, A.-P.
  Li, Z.~Jiang, E.~H. Conrad, C.~Berger, C.~Tegenkamp, W.~A. de~Heer,
  Exceptional ballistic transport in epitaxial graphene nanoribbons, Nature 506
  (2014) 349.
\newblock \href {http://dx.doi.org/10.1038/nature12952}
  {\path{doi:10.1038/nature12952}}.

\bibitem{Duffy2016}
J.~Duffy, J.~Lawlor, C.~Lewenkopf, M.~S. Ferreira,
  \href{https://link.aps.org/doi/10.1103/PhysRevB.94.045417}{Impurity
  invisibility in graphene: Symmetry guidelines for the design of efficient
  sensors}, Phys. Rev. B 94 (2016) 045417.
\newblock \href {http://dx.doi.org/10.1103/PhysRevB.94.045417}
  {\path{doi:10.1103/PhysRevB.94.045417}}.
\newline\urlprefix\url{https://link.aps.org/doi/10.1103/PhysRevB.94.045417}

\bibitem{Ridolfi2017}
E.~Ridolfi, L.~R.~F. Lima, E.~R. Mucciolo, C.~H. Lewenkopf,
  \href{https://link.aps.org/doi/10.1103/PhysRevB.95.035430}{Electronic
  transport in disordered {M}o{S}$_{2}$ nanoribbons}, Phys. Rev. B 95 (2017)
  035430.
\newblock \href {http://dx.doi.org/10.1103/PhysRevB.95.035430}
  {\path{doi:10.1103/PhysRevB.95.035430}}.
\newline\urlprefix\url{https://link.aps.org/doi/10.1103/PhysRevB.95.035430}

\bibitem{neto2009electronic}
A.~H. Castro~Neto, F.~Guinea, N.~M.~R. Peres, K.~S. Novoselov, A.~K. Geim,
  \href{https://link.aps.org/doi/10.1103/RevModPhys.81.109}{The electronic
  properties of graphene}, Rev. Mod. Phys. 81 (2009) 109--162.
\newblock \href {http://dx.doi.org/10.1103/RevModPhys.81.109}
  {\path{doi:10.1103/RevModPhys.81.109}}.
\newline\urlprefix\url{https://link.aps.org/doi/10.1103/RevModPhys.81.109}

\bibitem{Duarte_2021}
F.~R. Duarte, S.~Mukim, A.~Molina-S{\'{a}}nchez, T.~G. Rappoport, M.~S.
  Ferreira, \href{https://doi.org/10.1088/1367-2630/ac10cf}{Decoding the {DC}
  and optical conductivities of disordered {MoS}2 films: an inverse problem},
  New Journal of Physics 23~(7) (2021) 073035.
\newblock \href {http://dx.doi.org/10.1088/1367-2630/ac10cf}
  {\path{doi:10.1088/1367-2630/ac10cf}}.
\newline\urlprefix\url{https://doi.org/10.1088/1367-2630/ac10cf}

\bibitem{Lima2018}
L.~R.~F. Lima, A.~Dusko, C.~Lewenkopf, {Efficient method for computing the
  electronic transport properties of a multiterminal system}, Phys. Rev. B
  97~(16) (2018) 165405.
\newblock \href {http://dx.doi.org/10.1103/PhysRevB.97.165405}
  {\path{doi:10.1103/PhysRevB.97.165405}}.

\bibitem{Lima2021}
L.~R.~F. Lima, C.~Lewenkopf, Local equilibrium charge and spin currents in
  two-dimensional topological systems (2021).
\newblock \href {http://arxiv.org/abs/2104.07206} {\path{arXiv:2104.07206}}.

\bibitem{MeirWingreen1992}
Y.~Meir, N.~S. Wingreen, {Landauer formula for the current through an
  interacting electron region}, Phys. Rev. Lett. 68 (1992) 2512.
\newblock \href {http://dx.doi.org/10.1103/PhysRevLett.68.2512}
  {\path{doi:10.1103/PhysRevLett.68.2512}}.

\bibitem{lewenkopf2013recursive}
C.~H. Lewenkopf, E.~R. Mucciolo, {The recursive Green's function method for
  graphene}, J. Comput. Electron. 12 (2013) 203.
\newblock \href {http://dx.doi.org/10.1007/s10825-013-0458-7}
  {\path{doi:10.1007/s10825-013-0458-7}}.

\bibitem{lee1985universal}
P.~A. Lee, A.~D. Stone,
  \href{https://link.aps.org/doi/10.1103/PhysRevLett.55.1622}{Universal
  conductance fluctuations in metals}, Phys. Rev. Lett. 55 (1985) 1622--1625.
\newblock \href {http://dx.doi.org/10.1103/PhysRevLett.55.1622}
  {\path{doi:10.1103/PhysRevLett.55.1622}}.
\newline\urlprefix\url{https://link.aps.org/doi/10.1103/PhysRevLett.55.1622}

\bibitem{Beenakker1997}
C.~W.~J. Beenakker,
  \href{https://link.aps.org/doi/10.1103/RevModPhys.69.731}{Random-matrix
  theory of quantum transport}, Rev. Mod. Phys. 69 (1997) 731--808.
\newblock \href {http://dx.doi.org/10.1103/RevModPhys.69.731}
  {\path{doi:10.1103/RevModPhys.69.731}}.
\newline\urlprefix\url{https://link.aps.org/doi/10.1103/RevModPhys.69.731}

\end{thebibliography}


\begin{thebibliography}{1}
\providecommand{\natexlab}[1]{#1}
\providecommand{\url}[1]{\texttt{#1}}
\expandafter\ifx\csname urlstyle\endcsname\relax
  \providecommand{\doi}[1]{doi: #1}\else
  \providecommand{\doi}{doi: \begingroup \urlstyle{rm}\Url}\fi

\bibitem[Mukim et~al.(2020)Mukim, Amorim, Rocha, Muniz, Lewenkopf, and
  Ferreira]{shardul}
S.~Mukim, F.~P. Amorim, A.~R. Rocha, R.~B. Muniz, C.~Lewenkopf, and M.~S.
  Ferreira.
\newblock Disorder information from conductance: A quantum inverse problem.
\newblock \emph{Phys. Rev. B}, 102:\penalty0 075409, Aug 2020.
\newblock \doi{10.1103/PhysRevB.102.075409}.
\newblock URL \url{https://link.aps.org/doi/10.1103/PhysRevB.102.075409}.

\end{thebibliography}

\end{document}

%\preprint{APS/123-QED}

%\title{Disorder information from conductance fingerprints: a quantum inverse problem}% Force line breaks with \\
\title{Spatial mapping of disordered 2D systems: the conductance Sudoku\\
Supplemental Material
}% Force line breaks with \\

%\author{S. Mukim$^{1}$, C. Lewenkopf and M. S. Ferreira$^{1,2}$}
%\address{$^1$School of Physics, Trinity College Dublin, Dublin 2, Ireland}
% \address{$^2$Centre for Research on Adaptive Nanostructures and Nanodevices (CRANN) \& Advanced Materials and Bioengineering Research (AMBER) Centre, Trinity College Dublin, Dublin 2, Ireland}
% \address{$^3$ Instituto de F\'{\i}sica Te\'orica, S\~ao Paulo State University, 01140-070, S\~ao Paulo, Brazil}
% \address{$^4$ Instituto de F\'{\i}sica, Universidade Federal Fluminense, 24210-346 Niterói, Brazil}

\author{S. Mukim}
\address{School of Physics, Trinity College Dublin, Dublin 2, Ireland}
\author{C. Lewenkopf}
\address{Instituto de F\'{\i}sica, Universidade Federal Fluminense, 24210-346 Niterói, Brazil}
\author{M. S. Ferreira}
\address{School of Physics, Trinity College Dublin, Dublin 2, Ireland}
\address{Centre for Research on Adaptive Nanostructures and Nanodevices (CRANN) \& Advanced Materials and Bioengineering Research (AMBER) Centre, Trinity College Dublin, Dublin 2, Ireland}

\date{\today}

%\begin{abstract}

%\end{abstract}
%\begin{keywords}
%editorials, notices, miscellaneous
%\end{keywords}
%display desired
\maketitle
%\tableofcontents
\section{Formulation of the inversion method for an arbitrary number of terminals and partitions}

\begin{figure}
    \centering
    \begin{subfigure}[b]{0.45\columnwidth}
        \centering
        \includegraphics[width = \columnwidth]{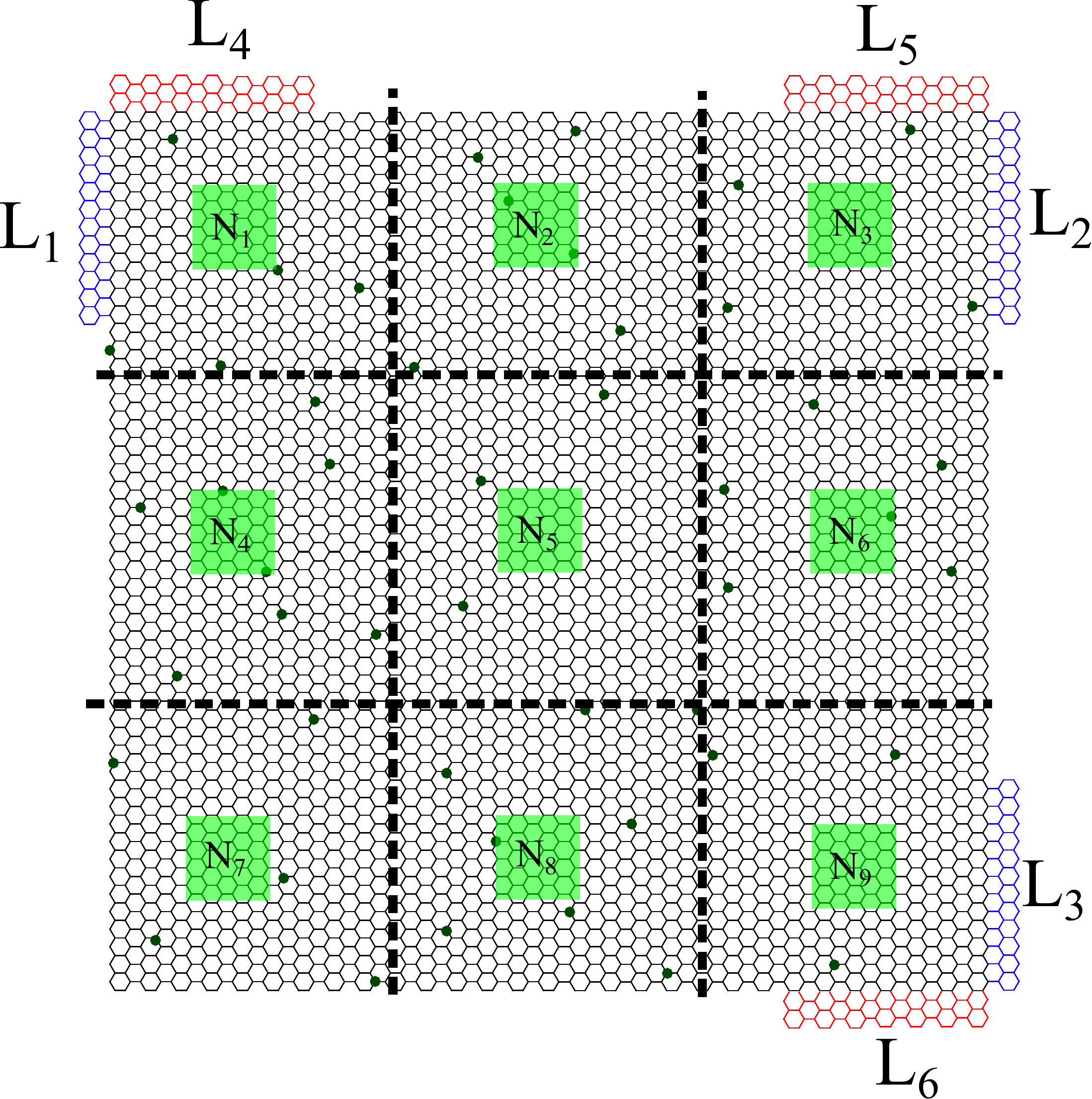}
        \caption{$3\times3$}
    \end{subfigure}
    \hfill
    \begin{subfigure}[b]{0.45\columnwidth}
        \centering
        \includegraphics[width=\columnwidth]{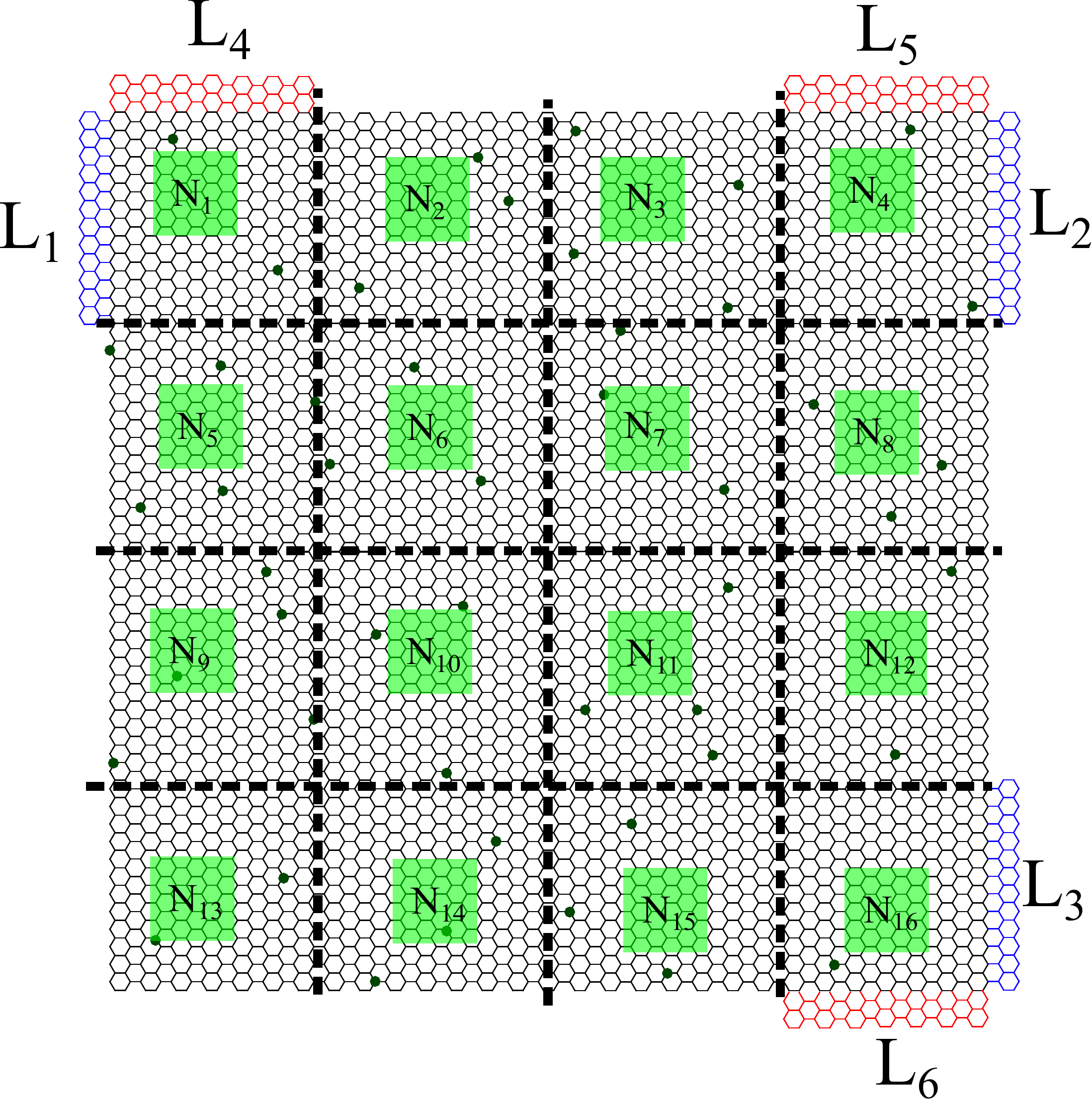}
       \caption{$4\times4$}
    \end{subfigure}
    \caption{A schematic representation of graphene flake embedded with substitutional disorder with a six electrode setup labelled as $L_{i}$, with $i$ running from $1$ to $6$. Dashed lines separate the flake into 9 and 16 regions and $N_j$ represents the number of impurities contained in each of these cells. }
    \label{fig1sm}
\end{figure}

Flexibility of the inversion procedure allows more than one approach to extract spatial information of the disordered components inside the flake. In the main manuscript the graphene flake is resolved into four quarters making use of a 6 electrode setup. However this is by no means the only arrangement used to map impurities in the graphene flake.
%\sout{Using shorter electrodes, it is possible to resolve the impurity distribution into a larger number of smaller cells.}
Fig.~\ref{fig1sm}(a) and \ref{fig1sm}(b) show a schematic representation of the graphene flake connected with six electrodes, together with imaginary dashed (blue) lines resolving the graphene flake into 9 and 16 cells, respectively. In main manuscript a system of four linear equations is defined to resolve the flake in quarter cells. To resolve the flake in 9 or 16 cells we need to define a system of linear equations which can be written in the form 
\begin{equation}
    AX - B = \lambda\,\,,
    \label{sysoflinear}
\end{equation}
where $A$ is a square matrix and the other elements are all single-column vectors whose size matches the number of cells. For example, in the case of 9 cells the equation takes the form of Eq.(~\ref{sysoflinear}) with $\lambda = 0$. Matrix A contains information about regions chosen to define 9 misfit functions such that ${\rm det} A \neq 0$. Variables $N_{i}$, with $i$ running from 1 to 9, indicate the impurity occupation of the individual cells shown in Fig.~\ref{fig1sm}(a) and $M_{i}$ is the occupation of corresponding region obtained from the  inversion. Note that we have flexibility on how we choose to interrogate the different parts of the flake, which means that the form of matrix $A$ is not unique but depends on that exact choice. Eq.~\ref{sys33} shows one such choice of a system of linear equations for the given arrangement of electrodes. 
\begin{equation}
\left(
\begin{array}{ccccccccc}
 1 & 1 & 1 & 0 & 0 & 0 & 0 & 0 & 0 \\
 0 & 0 & 0 & 1 & 1 & 1 & 0 & 0 & 0 \\
 1 & 1 & 1 & 1 & 1 & 1 & 1 & 1 & 1 \\
 1 & 0 & 0 & 1 & 0 & 0 & 1 & 0 & 0 \\
 0 & 1 & 0 & 0 & 1 & 0 & 0 & 1 & 0 \\
 1 & 1 & 0 & 1 & 1 & 0 & 0 & 0 & 1 \\
 1 & 0 & 0 & 1 & 0 & 0 & 0 & 1 & 1 \\
 1 & 0 & 0 & 0 & 1 & 1 & 0 & 1 & 1 \\
 1 & 1 & 0 & 0 & 0 & 1 & 0 & 0 & 1 \\
\end{array}
\right).\left(
\begin{array}{c}
 N_1 \\
 N_2 \\
 N_3 \\
 N_4 \\
 N_5 \\
 N_6 \\
 N_7 \\
 N_8 \\
 N_9 \\
\end{array}
\right)=\left(
\begin{array}{c}
 M_1 \\
 M_2 \\
 M_3 \\
 M_4 \\
 M_5 \\
 M_6 \\
 M_7 \\
 M_8 \\
 M_9 \\
\end{array}
\right)
\label{sys33}
\end{equation}
For example, the top-row identity from the matrix equation above reads that $N_1 + N_2 +N_3 = M_1$. By minimising a misfit function defined in terms of the variable $M_1$, we obtain the exact number of impurities in the top row of the flake, as shown in Fig.~\ref{fig1sm}(a). The same must be done with other regions such that we find the values of all $M_{i}$ in Eq.(\ref{sys33}). Analogously to the four-cell case, the inversion is carried out with a few different readings. In this case, $M_1$, $M_2$ and $M_3$ are found from ${\cal{T}}_{1,2}$; $M_4$, $M_5$ and $M_6$ are found from ${\cal{T}}_{5,6}$;  $M_7$ and $M_8$ are found from ${\cal{T}}_{1,3}$; finally $M_9$ is obtained from ${\cal{T}}_{4,6}$.

The remaining task is to solve the sudoku-style puzzle of identifying the impurity numbers $N_i$ of individual cells constrained to match the values of $M_{i}$.   
%Solving Eq.(~\ref{sys33}) is an easy task for obtained values of $M_{i}$. 
We obviously must account for the possibility of error in the inversion procedure which would lead to the incorrect mapping of the disorder concentration. This can be alleviated by obtaining more information about the flake and availing of more misfit functions, which will then serve as extra constraints in Eq.(\ref{sysoflinear}). Constrained optimization methods like Lagrange multiplier or differential evolution are very well established procedures and are suitable to be used here to find the values of $N_j$.   
%\begin{table}[!h]
%    \centering
%    \begin{tabularx}{0.45\textwidth} {| > {\centering\arraybackslash} X | > {\centering\arraybackslash} X| > {\centering\arraybackslash} X |}
%    \hline
%    7 (\textcolor {blue} {6}) & 8 (\textcolor {blue} {7}) & 5 (\textcolor {blue} {6}) \\
%    \hline
%    8 (\textcolor {blue} {7}) & 5 (\textcolor {blue} {4}) & 7 (\textcolor {blue} {6}) \\
%    \hline
%    4 (\textcolor {blue} {5}) & 5 (\textcolor {blue} {6}) & 3 (\textcolor {blue} {3}) \\
%    \hline
%    \end{tabularx}
%    \label{tablesm}
%\end{table}

\begin{figure}[!h]
    \centering
    \includegraphics[width=0.5\columnwidth]{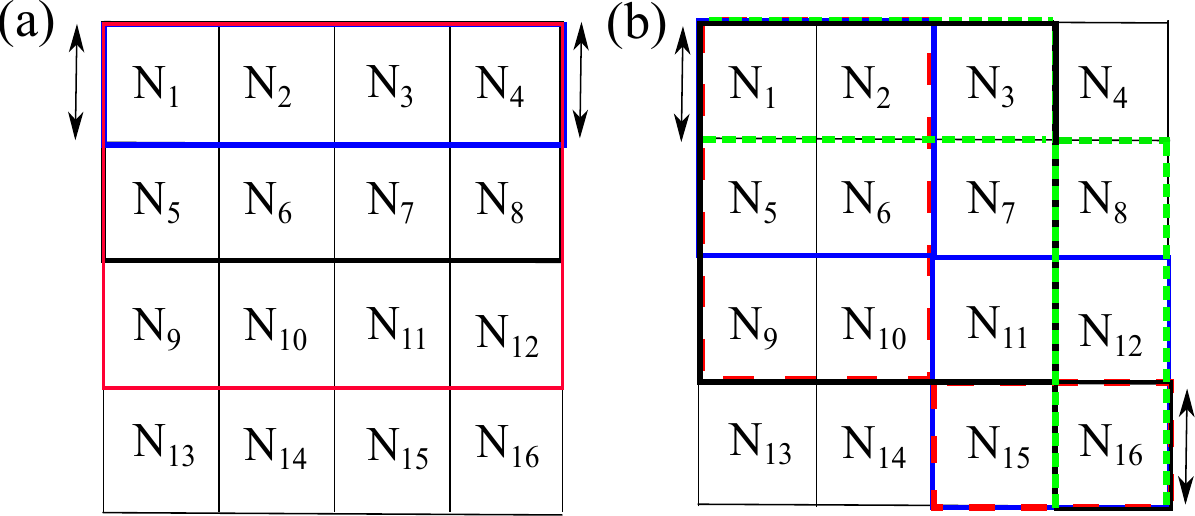}
    \caption{Flake is resolved into 16 cells with $N_{i}$ running from 1 to 16 gives occupation of each cell of the flake.(a) Misfit function and corresponding CA is defined for the regions enclosed in blue, black and red lines for input signal generated from leads position presented in the form of double arrow lines. (b) For the leads positioned diagonally opposite of the flake dashed red green lines shows non-trivial consideration of region for the inversion.}
    \label{reference_16}
\end{figure}

Left half of Table I in the main manuscript displays the numbers found for the nine cells together with the real occupation (in brackets). The agreement between the numbers is very good and the discrepancy is not more than a single impurity. 
%In the following section we explain the inversion Sudoku problem for 16 quarters in details. 
%\section{$4 \times 4$}
%In order to construct the spatial map of disorder, information is obtained from the transmission spectra generated by the set of red and green electrodes. 
%The corresponding system of linear of equations can be written in following form 
%In the case of nine quarters as seen in Fig. \ref{fig1sm}, matrix A is of size $9 \times 9$ and contains relevant information from different quarters of the flake that defines misfit functions. Note that misfit functions are defined such that $|A| \neq 0$. Matrix B has the information obtained from misfit functions. 
%For example using input ${\cal{T}}_{1,3}$ misfit function can be defined as $\chi(N_{1}+N_{2}+N_{4}+N_{5}+N_{9})$. This misfit function yields occupation of disorders in the region constrained by the variables.
%In the main manuscript we present a way to interrogate a system with half length electrodes in 4 blocks. 
%\begin{figure}[!h]
 %   \centering
%    \includegraphics[width=0.5\columnwidth]{fig_2_sm.pdf}
%    \caption{A schematic representation of graphene flake embedded with substitutional disorders. Electrodes labelled as $L_{i}$ $i$ running from $1$ to $6$. Dashed lines defines the flake in 16 different regions labelled as $N_{1}$ to $N_{16}$.}
%    \label{fig3sm}
%\end{figure}
In the case of sixteen cells, the dashed lines of Fig.~\ref{fig1sm} delineate them together with the corresponding occupation numbers $N_{i}$, with $i$ running from $1$ to $16$. Analogously to the previous cases considered and in line with Eq.(\ref{sysoflinear}), we must select regions based on which the misfit function will be calculated. In the case of four quarters in the main manuscript, ${\cal{T}}_{1,2}$ served as input to the misfit function in order to extract information about top half of the flake. The same method can be implemented to obtain occupation of each of the four rows. We define such misfit function as $\chi_{1,2}(N_{a},N_{b},N_{c},N_{d})$ where $N_{a}$ is the occupation of disorders in the first row, $N_{b}$ second row and so on as shown in Fig.~\ref{reference_16}(a) with blue, black and red lines contouring respective area of the flake. Note that occupation in rows is constrained by $N_{a}+N_{b}+N_{c}+N_{d}=N$. 

Similarly, misfit function defined using the input in form of ${\cal{T}}_{5,6}$ yields total occupation ($N_e$, $N_f$, $N_g$, and $N_h$) in each column of the flake in Fig.~\ref{fig1sm}(b). The remaining equations needed to complete the system of linear equations come from the diagonal misfit functions. Diagonal misfit functions are defined using input transmission spectra from leads which are diagonally opposite to each other, namely ${L}_{1,3}$ and  ${L}_{4,6}$. 
%We now have 4 equations with which flake is resolved in 4 rows from single input conductance ${\cal {T}}_{1,2}$. 
%In similar fashion we can extract information about the 4 columns by defining misfit function $\chi_{4,6}(N_{e},N_{f},N_{g},N_{h})$, here $N_{e}$ gives information about the occupation in the first column and so on.
%However this information is not sufficient to accurately map the disorders inside the flake. 
\begin{figure}[!h]
    \centering
    \includegraphics[width=0.5\columnwidth]{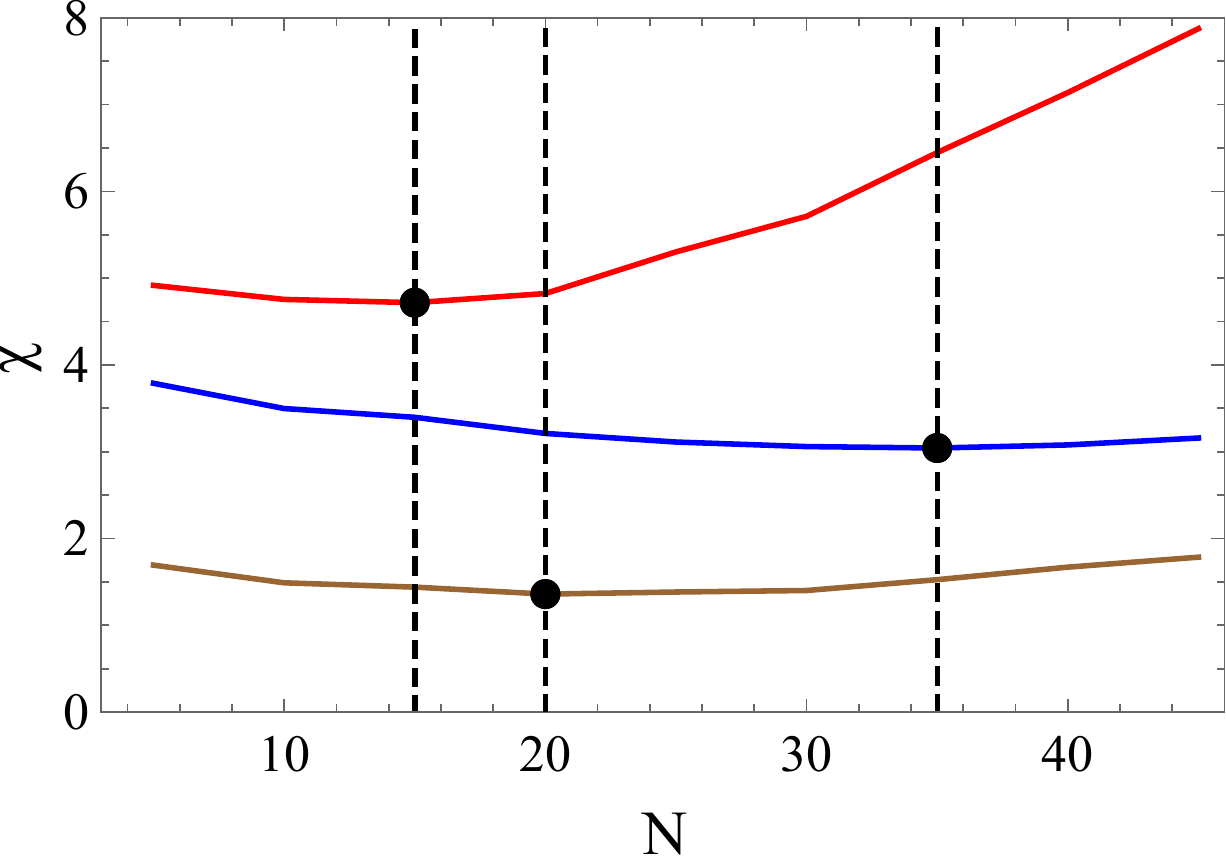}
    \caption{Misfit function in blue defined for region enclosed in the thick black line in Fig.~\ref{reference_16}(b) shows minimum at $N=35$ which closely coincides with $N=37$ actual number of disorders in region. Brown and red misfit curves locate the occupation of regions enclosed by dashed red line and dashed green line of Fig.~\ref{reference_16}, respectively. }
    \label{misfitdiag}
\end{figure}
We need to define a system of sixteen linear equations to obtain the occupation of each one of the cells. We have seven equations coming from ${L}_{1,3}$ and ${L}_{4,6}$. Two extra input signals from diagonally arranged leads will provide the remaining equations, namely ${\cal T}_{1,3}$ and ${\cal T}_{4,6}$. 
%Corresponding misfit function deals with diagonal quarters as shown in Fig.~\ref{reference_16}(b) as  variables. 
Using ${\cal T}_{1,3}$ as input the flake is scanned in four different regions as shown in Fig.~\ref{reference_16} contoured by dashed lines in black, green, red and blue colours. %In main manuscript Fig.2(b) shows the misfit function defined for the region contoured by blue color shows occupation as $N_{D}=20$. 

In Fig.~\ref{misfitdiag}, three misfit functions are plotted as a function of number of impurities. Misfit function in red is defined with variables enclosed by dashed red line shown in Fig.~\ref{reference_16}(b). Misfit function in blue is defined for region enclosed by black lines in the schematic representation seen in Fig.~\ref{reference_16}(b). It is worth mentioning that the misfit function is minimum at $N=35$ and when compared with the actual distribution of impurities in the parent configuration the discrepancy does not exceed two impurities. Finally, the misfit function defined using the input signal ${\cal T}_{4,6}$  completes the set of equations necessary to generate the system of linear equations. It is important to note that this is by no means the only way to resolve the device in sixteen cells. Different arrangement of leads may lead to different ways of mapping the impurities. Following our definition and choice for how the system was probed, the matrix A of Eq.(\ref{sysoflinear}) is given by
\begin{equation}
 A=\left(
\begin{array}{c c c c c c c c c c c c c c c c}
 1 & 1 & 1 & 1 & 0 & 0 & 0 & 0 & 0 & 0 & 0 & 0 & 0 & 0 & 0 & 0 \\
 0 & 0 & 0 & 0 & 1 & 1 & 1 & 1 & 0 & 0 & 0 & 0 & 0 & 0 & 0 & 0 \\
 0 & 0 & 0 & 0 & 0 & 0 & 0 & 0 & 1 & 1 & 1 & 1 & 0 & 0 & 0 & 0 \\
 0 & 0 & 0 & 0 & 0 & 0 & 0 & 0 & 0 & 0 & 0 & 0 & 1 & 1 & 1 & 1 \\
 1 & 0 & 0 & 0 & 1 & 0 & 0 & 0 & 1 & 0 & 0 & 0 & 1 & 0 & 0 & 0 \\
 0 & 1 & 0 & 0 & 0 & 1 & 0 & 0 & 0 & 1 & 0 & 0 & 0 & 1 & 0 & 0 \\
 0 & 0 & 1 & 0 & 0 & 0 & 1 & 0 & 0 & 0 & 1 & 0 & 0 & 0 & 1 & 0 \\
 1 & 1 & 0 & 0 & 1 & 1 & 0 & 0 & 0 & 0 & 1 & 1 & 0 & 0 & 1 & 1 \\
 1 & 1 & 1 & 0 & 1 & 1 & 1 & 0 & 1 & 1 & 1 & 0 & 0 & 0 & 0 & 1 \\
 1 & 0 & 0 & 0 & 0 & 1 & 1 & 1 & 0 & 1 & 1 & 1 & 0 & 1 & 1 & 1 \\
 1 & 1 & 1 & 0 & 1 & 1 & 1 & 0 & 0 & 0 & 0 & 1 & 0 & 0 & 0 & 1 \\
 1 & 1 & 0 & 0 & 1 & 1 & 0 & 0 & 1 & 1 & 0 & 0 & 0 & 0 & 1 & 1 \\
 1 & 1 & 0 & 0 & 0 & 0 & 1 & 1 & 0 & 0 & 1 & 1 & 0 & 0 & 1 & 1 \\
 1 & 0 & 0 & 0 & 1 & 0 & 0 & 0 & 0 & 1 & 1 & 1 & 0 & 1 & 1 & 1 \\
 1 & 1 & 1 & 0 & 0 & 0 & 0 & 1 & 0 & 0 & 0 & 1 & 0 & 0 & 0 & 1 \\
 1 & 0 & 0 & 0 & 1 & 0 & 0 & 0 & 1 & 0 & 0 & 0 & 0 & 1 & 1 & 1 \\
\end{array}
\right)
\label{16mat}
\end{equation}

%\begin{table}[!h]
%    \centering
%    \begin{tabularx}{0.45  \textwidth} { 
%  | >{\centering\arraybackslash}X 
%  | >{\centering\arraybackslash}X 
%  | >{\centering\arraybackslash}X 
%  | >{\centering\arraybackslash}X |}
%    \hline
%     0 (\textcolor{blue}{2}) & 2 (\textcolor{blue}{3}) & 5 (\textcolor{blue}{4}) & 3 (\textcolor{blue}{4})\\
%    \hline
%     7 (\textcolor{blue}{5}) & 4 (\textcolor{blue}{3}) & 1 (\textcolor{blue}{2}) & 3 (\textcolor{blue}{3})\\
%     \hline
%     5 (\textcolor{blue}{5}) & 2 (\textcolor{blue}{3}) & 6 (\textcolor{blue}{4}) & 2 (\textcolor{blue}{2}) \\
%      \hline
%     3 (\textcolor{blue}{2}) & 2 (\textcolor{blue}{3}) & 3 (\textcolor{blue}{3}) & 2 (\textcolor{blue}{2}) \\
%    \hline
%    \end{tabularx}
%    \label{tab:16}
%\end{table}

After solving this system of linear equations we can finally map the impurity distribution across the flake and this is shown on the right part of Table I in the main manuscript. The table shows the real occupation of impurities for all sixteen cells together with the sudoku-style solutions.

\section{Role of the system-electrode contact geometry}
\begin{figure}[!h]
    \centering
    \begin{subfigure}[b]{0.49\textwidth}
    \centering
    \includegraphics[width=0.850\columnwidth]{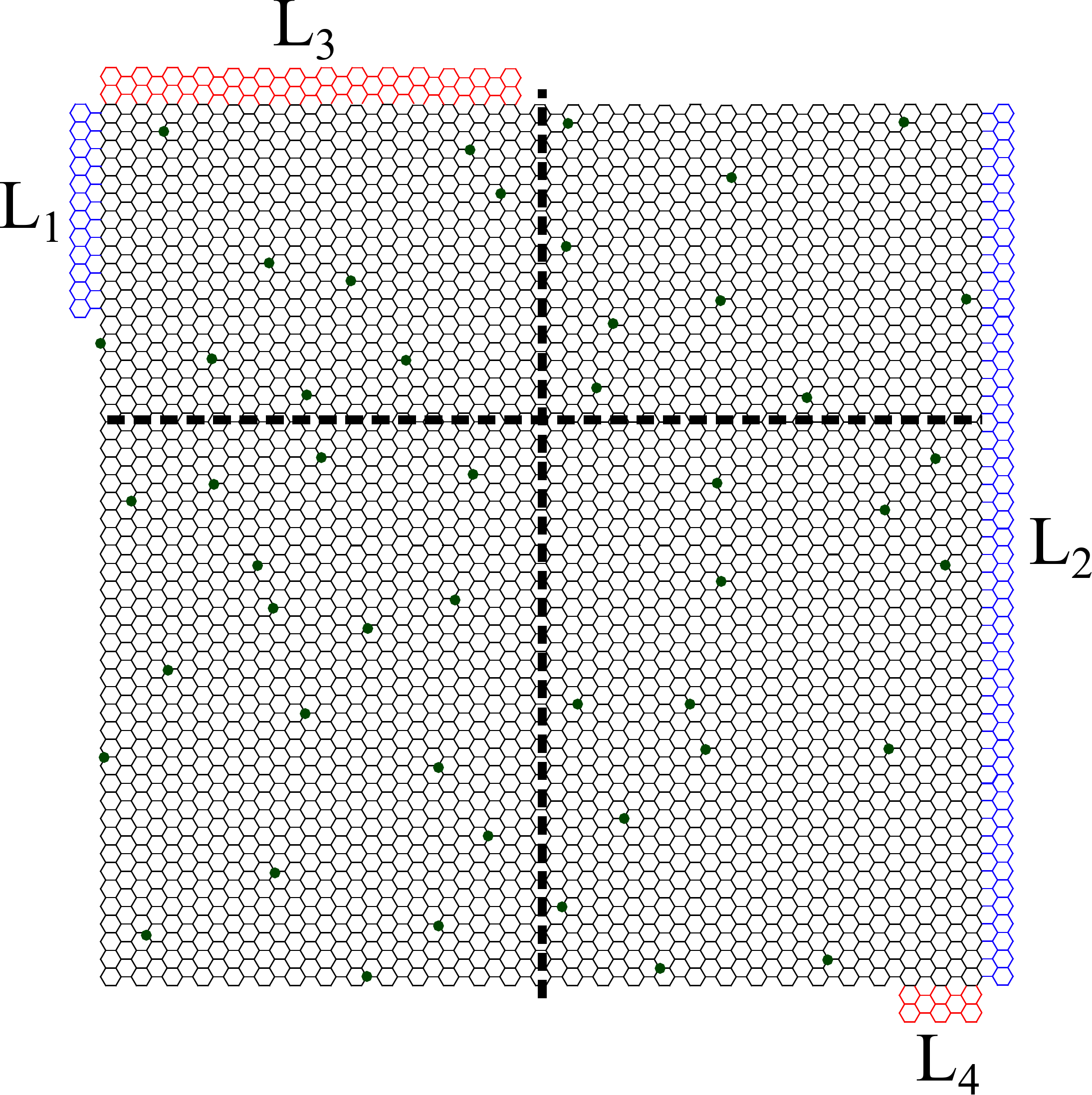}
    \label{schematic_hybrid}
    \caption{}
    \end{subfigure}
    \begin{subfigure}[b]{0.49\textwidth}
    \centering
    \includegraphics[width=0.850\columnwidth]{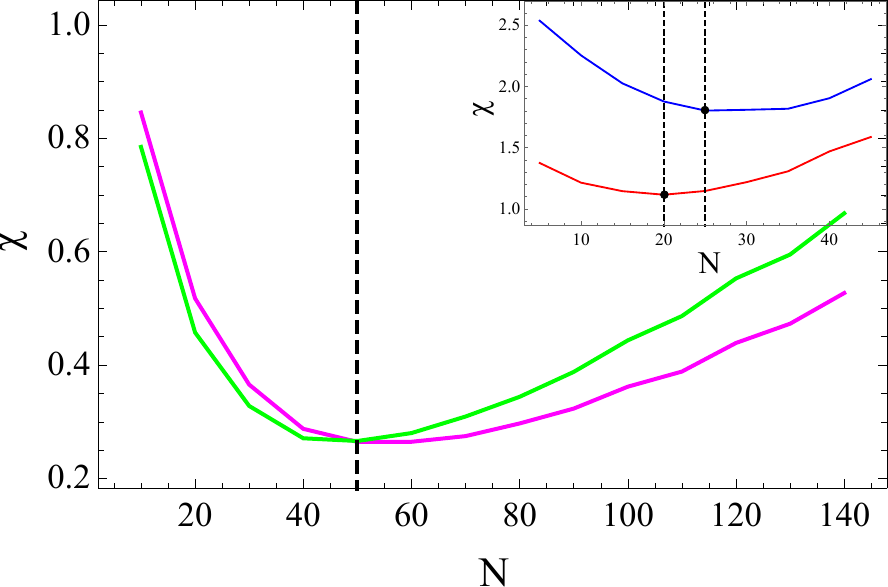}
    \label{hybrid_misfit}
    \caption{}
    \end{subfigure}
    \caption{(a) A schematic presentation of the hybrid approach where one lead spans entire width of the and other electrodes are of different lengths. (b) Misfit function plotted in blue color is defined using input signal ${\cal T}_{3,4}$ scans left half of the device. Misfit function defined using  ${\cal T}_{1,2}$ yields occupation of row constrained by pink leads and dashed blue lines . }
    \label{hybrid_appr}
\end{figure}

In the first part of this SM we showed how flexible this inversion technique is to the extent that we may obtain the spatial distribution of impurities on a 2D flake in a few different ways. With this in mind, we now illustrate yet another way in which the impurity distribution can be spatially mapped, this time using electrodes of different lengths. In the main manuscript we considered a 6 electrode setup. We considered the case in which both injecting and extracting electrodes were of similar in size. We  now relax this constraint and assume that electrodes have different dimensions, which we refer to as the hybrid approach. A schematic representation of this can be seen in Fig.\ref{hybrid_appr} where one lead spans the entire width of the flake whereas the other is either half, a quarter of the length of its counterpart or even shorter than the quarter length, as size of the electrode is not a critical factor in the IP. It is important to stress that the exact setups, either the one shown in Fig.~1 of the main manuscript or the one in Fig.\ref{hybrid_appr} here, are not unique regarding the positions and respective sizes of the electrodes being used. The key here is to have them in a non-symmetric disposition. In other words, the inversion success is not sensitive to the precise electrodes setup probing the conductance matrix. This makes the experimental realisation far less challenging than if it depended on the exact and precise placements of the electrodes. Here we follow similar steps to the ones taken previously and find exactly the same results as the ones found in Fig.1 of the main manuscript. In fact, Fig.~\ref{hybrid_appr}(b) displays the  misfit functions plotted as a function of impurity numbers in the entire flake using leads in magenta and green colors. Although the functions are different, both display a distinctive minimum at the same value of $N=50$. In the inset another two misfit functions are plotted with minima indicating the occupation of the top row (red) and left half (blue) of the flake indicated with the dashed lines of Fig.~\ref{hybrid_appr}.

%\begin{figure}[!h]
%    \centering
%    \includegraphics[width=0.5\columnwidth]{hybrid.pdf}
%    \caption{Misfit function plotted in blue color is defined using input signal ${\cal T}_{3,4}$ scans left half of the device. Misfit function defined using  ${\cal T}_{1,2}$ yields occupation of row constrained by pink leads and dashed blue lines . }
%    \label{hybrid_misfit}
%\end{figure}

\section{Generalised partitions}
\begin{figure}[!h]
    \centering
    \includegraphics[width=0.5\textwidth]{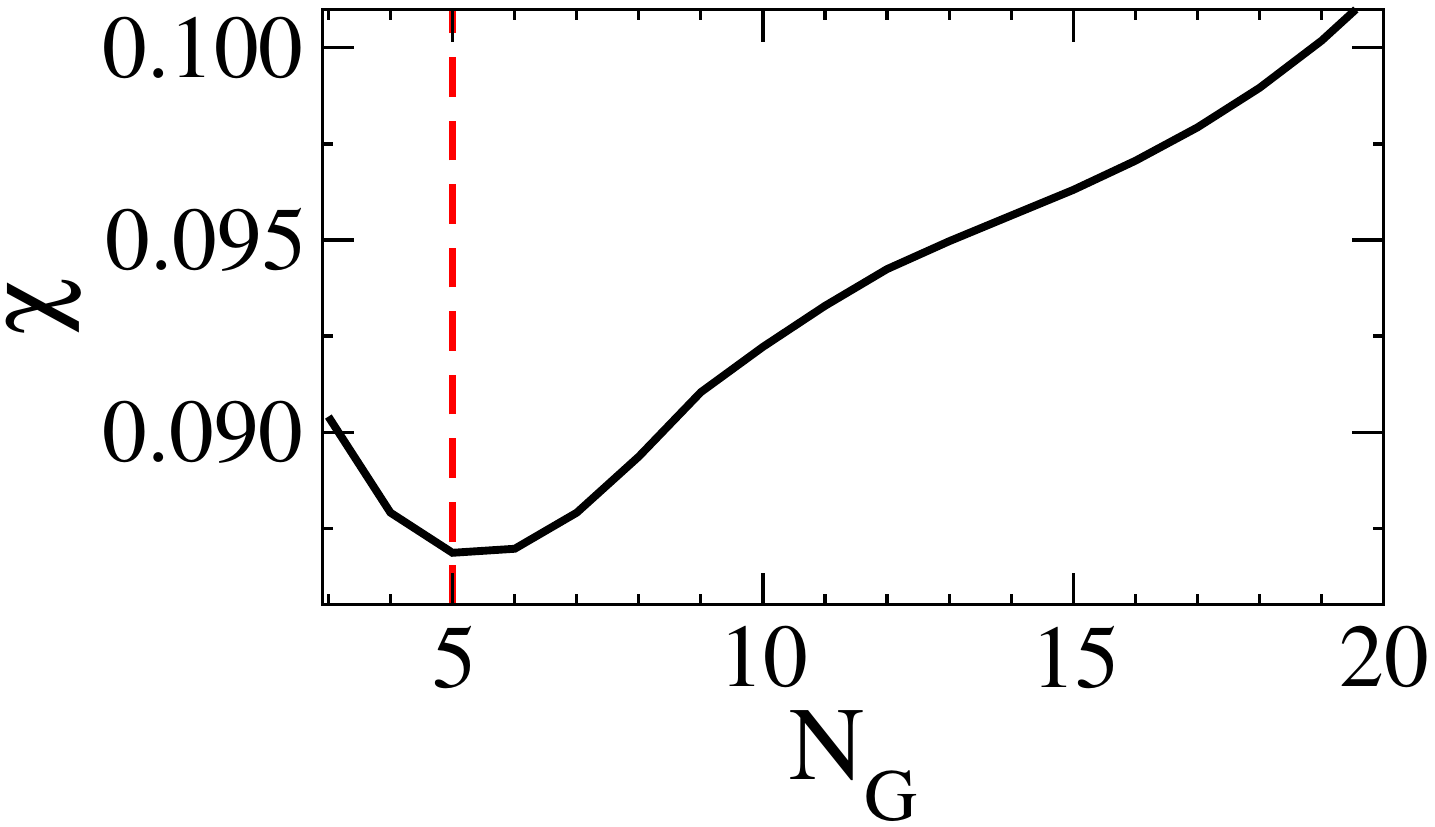}
    \caption{Misfit function defined for an arbitrary green region with $N_{G}$ as the impurity number for that region. Minimum at $N_{G} = 5 $ identifies the number of impurities for the green cell of Fig.3(c). The previously obtained input spectrum ${\cal{T}}_{1,3}$ was used to define the misfit function. }
    \label{MISFIT_GREEN}
\end{figure}

In the main manuscript and previous sections of the SM, we presented a simple methodology to map the disorder of the system in the Sudoku-style structure of the flake. We resolved the flake into 4, 9 and 16 cells. Suppose that we are interested in establishing the impurity concentration within a certain sector of the device that does not coincide with any of the cells defined previously. In that case, instead of carrying out a full sudoku-style mapping which would involve defining a whole range of other cells, we may avail of the information obtained for the 4, 9 and/or 16 cells as a constraint in the search for the occupation in the cell of interest. Consider for example the green cell highlighted in Fig.3(c) of the main manuscript. That cells does not match with any other previously used cells but it is possible to find the impurity number $N_G$ within that green cell by carrying out an additional CA calculation. This extra calculation is not particularly intense and simply involves a new misfit function that depends on $N_G$ as its key variable but uses the information already obtained as constraints. Assuming that the green cell lies within one of the four sectors defined in the $2 \times 2$ case, whose solution is known from a previous inversion to be $N_4$, the misfit function that captures the problem at hand is $\chi(N_{G},N, N_{4})$. Using $N$ and $N_4$ as constraints, we show in Fig.~\ref{MISFIT_GREEN} this misfit function that displays a clear minimum at $N_G = 5$, which closely matches the actual number seen in Fig.3(c). 
It is important to note that no additional input spectrum is needed to resolve the impurity concentration within an arbitrary region of the flake besides the readings already carried out.  

%\sout{however using this information we can investigate the flake in a more flexible way as shown in Fig.3(c). Shaded green region corresponds to an arbitrary region of the device, not defined in Fig.~\ref{fig1sm}. However we may still use the information provided by the results of 9 and 16 cells to focus on the  green cell seen in Fig.3(c).} \mauro{\sout{This calls for one extra CA calculation to generate one additional misfit function, namely $\chi(N_{G},N, N^{\alpha}_{4})$. $N_G$ is the impurity number in the green cell, $N$ is the total impurity number and}  }

%\sout{IP allows flexibility to define misfit function to obtain occupation of disorders in green region as $\chi(N_{G},N, N^{\alpha}_{4})$. Here in $N^\alpha(\alpha = 4, 9, 16)$ indicates the occupation obtained while resolving flake respective regions. In this particular case green region is overlaps regions defined by $N_{11}$,$N_{12}$,$N_{15}$ and $N_{16}$ of $4\times4$ mapped flake. In addition to that we already have information about the regions defined by $3\times 3$ mapping of the flake. Regions $N_{5}$,$N_{6}$, $N_{8}$ and $N_{9}$ in addition to the cells obtained from mapping of $4\times4$ helps to constrain while solving for the green region. Fig.~\ref{MISFIT_GREEN} shows the misfit plotted as a function of $N_{G}$ shows a minimum at $N_{G}= 5$ constrained by $N_{total}=50$, and Table I in main manuscript provides occupation of the cells overlapping green region.} 
 
\section{Scalability}

In the manuscript, we presented a six electrode setup in Fig.1(a) to resolve the device in different ways. A minimum of four input signals are required to resolve the flake into sixteen cells. As stated in the main manuscript, by increasing the number of cells the resolution accuracy drops. Table 2 in the main manuscript encapsulates how the accuracy changes as a function of resolution. Accuracy of the spatial mapping depends on three main parameters: (i) size of the system, (ii) impurity concentration and (iii) number of input readings. It can be shown that the error decreases with larger device sizes and with a reduced impurity concentration \cite{shardul}. Therefore, increasing our device size will make the accuracy values seen in Table 2 significantly better. For example, device sizes in the range of $\mu m$ are likely to present errors well below $20\%$ when divided into the same sixteen cells shown in the manuscript. While there are no methodology limitations to considering devices sizes in this range, the computational cost of carrying out a full inversion in a system of that size makes this task unfortunately very challenging at this point in time. However, by increasing the number of electrodes one can generate more input signals which in turn can provide additional information about the spatial distribution of the disorder resulting in the improvement of the accuracy, similarly to the case of generic reconstruction algorithms.

\bibliographystyle{abbrvnat}
\bibliography{mybib}

%\begin{figure}
%    \centering
%    \includegraphics[width=0.5\linewidth]{error_sudo.eps}
%    \caption{Error as a function of number of cells used to resolve the device. When resolved into 16 cells, each cell of dimensions approximately $13a*13a$ has an error of 0.12.}
%    \label{error}
%\end{figure}

%\begin{figure}[!h]
%    \centering
 %   \includegraphics[width= \columnwidth]{4_1_d_leads.pdf}
 %   \caption{Caption}
%    \label{fig:my_label}
%\end{figure}

%--------------------------------